\documentclass[a4paper,preprint,review,authoryear,1p,12pt]{elsarticle}

\usepackage{epsfig}

\usepackage{booktabs}       
\usepackage[table,xcdraw]{xcolor}

\usepackage{amssymb}
\usepackage{enumitem}

\usepackage{color}
\usepackage{graphicx}
\usepackage{subfigure}

\usepackage{amsmath}      
\usepackage{array}  
    
\usepackage[titletoc]{appendix}
\usepackage[authoryear]{natbib}

\usepackage{algorithmic} 
\usepackage{amsthm}
\usepackage[figuresright]{rotating}
\usepackage{subeqnarray}
\usepackage{cases}
\usepackage{makecell}
\usepackage{lineno}

\usepackage{mathrsfs}
\usepackage[style=long]{glossaries}
\usepackage{subcaption}
\usepackage[ruled,linesnumbered]{algorithm2e}

\biboptions{square,comma}

\journal{}

\usepackage[pdftex,bookmarksnumbered,colorlinks,linkcolor=red,anchorcolor=blue,citecolor=green]{hyperref}
\usepackage{geometry}
\graphicspath{{figures/}}

\usepackage[english]{babel}
\usepackage{amsthm}

\usepackage{changes}

\usepackage{nomencl}
\makenomenclature

\begin{document}

\begin{frontmatter}

\title{Reinforcement Learning-Driven Plant-Wide Refinery Planning Using Model Decomposition}

\author{Zhouchang Li,
        Runze Lin,
        Hongye Su,
        Lei Xie*}
\address{Zhejiang University, State Key Laboratory of Industrial Control Technology, Hangzhou 310027\\
         E-mail: leix@iipc.zju.edu.cn}
\markboth{Your Name(s)}{Reinforcement Learning-Driven Plant-Wide Refinery Planning Using Model Decomposition}

\begin{abstract}

\noindent 

In the era of smart manufacturing and Industry 4.0, the refining industry is evolving towards large-scale integration and flexible production systems. In response to these new demands, this paper presents a novel optimization framework for plant-wide refinery planning, integrating model decomposition with deep reinforcement learning. The approach decomposes the complex large-scale refinery optimization problem into manageable submodels, improving computational efficiency while preserving accuracy. A reinforcement learning-based pricing mechanism is introduced to generate pricing strategies for intermediate products, facilitating better coordination between submodels and enabling rapid responses to market changes. Three industrial case studies, covering both single-period and multi-period planning, demonstrate significant improvements in computational efficiency while ensuring refinery profitability.
\end{abstract}

\begin{keyword}

Large scale refinery planning, reinforcement learning, model decomposition, pricing strategy.
\end{keyword}

\end{frontmatter}

\section{Introduction}

Under the paradigm of smart manufacturing and Industry 4.0, the refining industry is evolving toward large-scale integration and flexible production systems \cite{bogle2017perspective,yuan2017smart}. Modern refineries must achieve economies of scale while maintaining operational adaptability to cope with dynamic market conditions. However, fierce industry competition and increasingly stringent environmental regulations have significantly reduced profit in the petrochemical industry \cite{kallrath2002planning,li2020development,olaizola2022refinery}. Uncertainties such as frequent fluctuations in the price of crude oil and petrochemical product also exacerbate the industry's challenging situation \cite{niu2020learning,awudu2013stochastic,shah2011petroleum}. These factors collectively necessitate the development of intelligent planning strategies to optimize refinery operations, ensuring profitability while meeting sustainability and compliance requirements.

Traditional refinery planning methods mainly involve constructing first-principles models of the full process flowsheet and applying mathematical programming techniques to identify optimal solutions. In earlier studies, researchers predominantly employed mixed-integer linear programming (MILP) models to formulate refinery production planning problems, primarily motivated by computational tractability and solution convenience \cite{pongsakdi2006financial,kuo2008application}. Although the MILP model provides tractable solutions, it often yields suboptimal or even infeasible results when handling complex refinery operations. Consequently, there is a growing trend to incorporate more accurate non-linear models during the optimization process to better capture the essence of refinery process, thus achieving a better balance between model precision and accuracy. Recent advances have demonstrated the superiority of non-linear modeling approach. Li et al. developed empirical nonlinear models for crude distillation unit (CDU) and fluid catalytic cracker (FCC) that explicitly account for crude oil properties, product yields, and quality parameters \cite{li2005integrating}. Menezes et al. subsequently enhanced the swing-cut model for atmospheric and vacuum distillation units, significantly improving the prediction accuracy of both product quantities and qualities \cite{menezes2013improved}. Building upon these advances, Li et al. proposed a tri-section CDU model incorporating binary variables, achieving modeling accuracy comparable to rigorous tray-to-tray approaches while maintaining computational efficiency in solving MINLP refinery planning problems with bilinear blending constraints \cite{li2021product}. Besides, some multi-period refinery planning mixed-integer nonlinear programming (MINLP) model was built based on a nonlinear fractionation index CDU model \cite{alattas2012refinery}, pour-point blending equations \cite{zhang2015multi} or distillation correlations with the Geddes fractionation index \cite{siamizade2019global}, enhancing both accuracy and profit.

To address the non-convex optimization challenges induced by nonlinearities, current approaches primarily focus on employing advanced mathematical decomposition algorithms and relaxation strategies. Non-convex generalized Benders decomposition was employed to address the uncertainty of crude oil quality \cite{yang2016integrated}. Castro et al. proposed a tightening piecewise McCormick relaxation method for bilinear problems \cite{castro2015tightening}, as well as a strategy for normalized multiparametric decomposition \cite{castro2016normalized}, achieving effective relaxation of mixed-integer bilinear problems. These two methodologies have been successfully implemented in refinery planning optimization by subsequent researchers \cite{castillo2017global,zhang2021refinery}, with computational experiments demonstrating their ability to achieve optimization results comparable to commercial solvers (e.g. BARON, ANTIGONE). Moreover, researchers have explored effective methods based on Lagrangian decomposition to tackle large-scale multi-period optimization problems \cite{uribe2023assessment,chu2015integrated,mouret2011new,yang2020integration}. Although stochastic optimization methods have also been explored \cite{vasant2012hybrid,pereira2020quantum,hou2015pareto,hou2016genetic}, such as genetic algorithms and hybrid tabu search algorithms, they often cannot guarantee an optimal solution as they tend to get trapped in local optima.

Despite demonstrated successes in refinery planning applications, current global optimization methods exhibit two fundamental limitations: the solution quality critically depends on cutting-plane selection strategies and scenario-specific modeling, and the required number of cutting planes grows exponentially with system scale. These challenges become particularly acute when addressing large-scale, fully integrated refinery systems with strong couplings, where traditional methods suffer from deterioration in computational efficiency, compromised solution quality, and reduced solution feasibility. This underscores the imperative for developing generalized optimization frameworks capable of automated constraint decomposition and integrated full-process coordination while maintaining polynomial-time complexity.

Recent advances in data availability and artificial intelligence present new opportunities to overcome these long-standing challenges \cite{daoutidis2024machine,decardi2024generative,jebreili2024optimization}. The accumulation of refinery operational data, coupled with breakthroughs in big data analytics and machine learning, has enabled data-driven approaches to complement traditional optimization paradigms. Notably, data-driven approaches, especially deep reinforcement learning (DRL), have demonstrated significant potential in process systems engineering \cite{lin2023accelerating,lin2024surrogate,chang2022controlling,ye2022coupling}, particularly for chemical engineering and refinery planning applications. Wang and Ning integrated distributed robust optimization with data-driven techniques, incorporating real operational data to improve model accuracy \cite{wang2022scenario,ning2019optimization}. Reinforcement learning has been successfully applied to multi-timescale multi-period decision-making by Shin \cite{shin2019multi}, enhancing strategy robustness and flexibility. Machine learning techniques, including principal component analysis and kernel density estimation, were widely adopted to construct uncertainty sets \cite{guevara2020machine,ning2018data,shang2017data,dai2020data}, while deep learning approaches like temporal graph convolutional networks (T-GCN) improved spatiotemporal price forecasting within optimization frameworks \cite{wang2024refinery}.
Data-driven methods have also proven effective in refinery process modeling. Surrogate models, such as radial basis function (RBF) neural networks, was developed for distillation processes \cite{lu2021surrogate}. A data-driven global optimization framework based on processing unit model parameters optimized using process datasets, showed significant profitability improvements in refinery planning \cite{li2016data}. Additionally, hybrid modeling combining historical data and centralized dynamics was utilized to optimize blending operations for FCC, catalytic reforming, and delayed coking units, reducing CO2 emissions while increasing profits \cite{li2024integrated}.

Although data-driven approaches have demonstrated promising results in certain refinery optimization applications, significant challenges remain in bridging the gap between theoretical validation and practical implementation. A primary limitation stems from the predominant focus of existing research on small-scale problems or restricted applications where data is merely used for parameter fitting, while real-world refinery involves the large-scale, multi-source, and highly coupled system which is inherently complex due to multivariate nature and strong interdependencies. Direct application of purely data-driven methods to such comprehensive planning problems often leads to suboptimal local solutions, training instability, and difficulties in handling nonlinear blending formula, as well as the intrinsic complexities arising from interactions among intermediate product streams, processing units, and storage systems. These limitations underscore the need for a hybrid methodological framework that effectively combines the strengths of data-driven techniques with mathematical optimization principles.

To address these challenges, this study proposes a plant-wide planning optimization framework integrating model decomposition with reinforcement learning. The core innovation lies in decomposing the large-scale refinery optimization problem into multiple simplified sub models, each corresponding to a specific operational segment of the refinery. This decomposition strategy systematically decouples interconnected variables and constraints, transforming the complex global problem into more tractable subproblems while maintaining computational efficiency. Furthermore, the framework incorporates a reinforcement learning-based pricing mechanism that synthesizes data patterns and topological features to develop comprehensive pricing strategies for decoupled sub models, particularly for intermediate products. The pricing strategy serves as a natural interface for hybrid algorithm design. By introducing price as signals, it can automatically coordinate material balances between sub-models, enabling rapid response to external disturbances through price adjustments. It effectively transforms complex globally-coupled optimization problems into distributed decision-making frameworks, making it particularly suitable for large-scale, coupled dynamic optimization problems. Furthermore, the price directly reflect the marginal values of intermediate products, providing management with intuitive decision-making indicators.

The proposed approach not only accommodates the nonlinear and dynamic characteristics of refinery operations but also leverages edge computing resources across distributed units to minimize communication overhead and latency. By synergistically combining these elements, the proposed framework maintains operational flexibility while pursuing globally optimal solutions that can adapt to evolving market conditions. To the best of our knowledge, there has been no research combining reinforcement learning and model decomposition methods to construct the optimization framework.

The main contributions of this paper can be summarized as follows:

\begin{itemize}
    \item A novel integrated framework combining model decomposition and reinforcement learning is proposed for plant-wide refinery planning optimization, which effectively handles large-scale, highly-coupled refinery planning problems.
    \item A generalized modeling framework is developed with systematic decomposition methodology and sub-model construction principles, ensuring both solution optimality and computational tractability.
    \item An innovative reinforcement learning-based pricing mechanism is designed to overcome local optima and training instability issues when applying RL directly to large-scale refinery planning, while better accommodating flexible production requirements. 
\end{itemize}

The structure of this article is organized as follows. The overall planning problem is introduced in Section \ref{Refinery planning description}. In Section \ref{Methodology}, we first describe the framework and rationale of our approach, which integrates pricing strategy based on DRL and model decomposition, and then present the necessary constraints for formulating the complete refinery planning model. This section also outlines the process of developing the pricing strategy. Section \ref{Case study} presents three industrial case studies where the proposed method is applied, comparing the results with those obtained from mathematical programming or RL, to demonstrate its validity. Finally, Section \ref{Conclusion} provides a summary of the study and highlights the key findings.

\section{Refinery planning description}\label{Refinery planning description}

\begin{figure}
    \centering
    \includegraphics[width=0.9\linewidth]{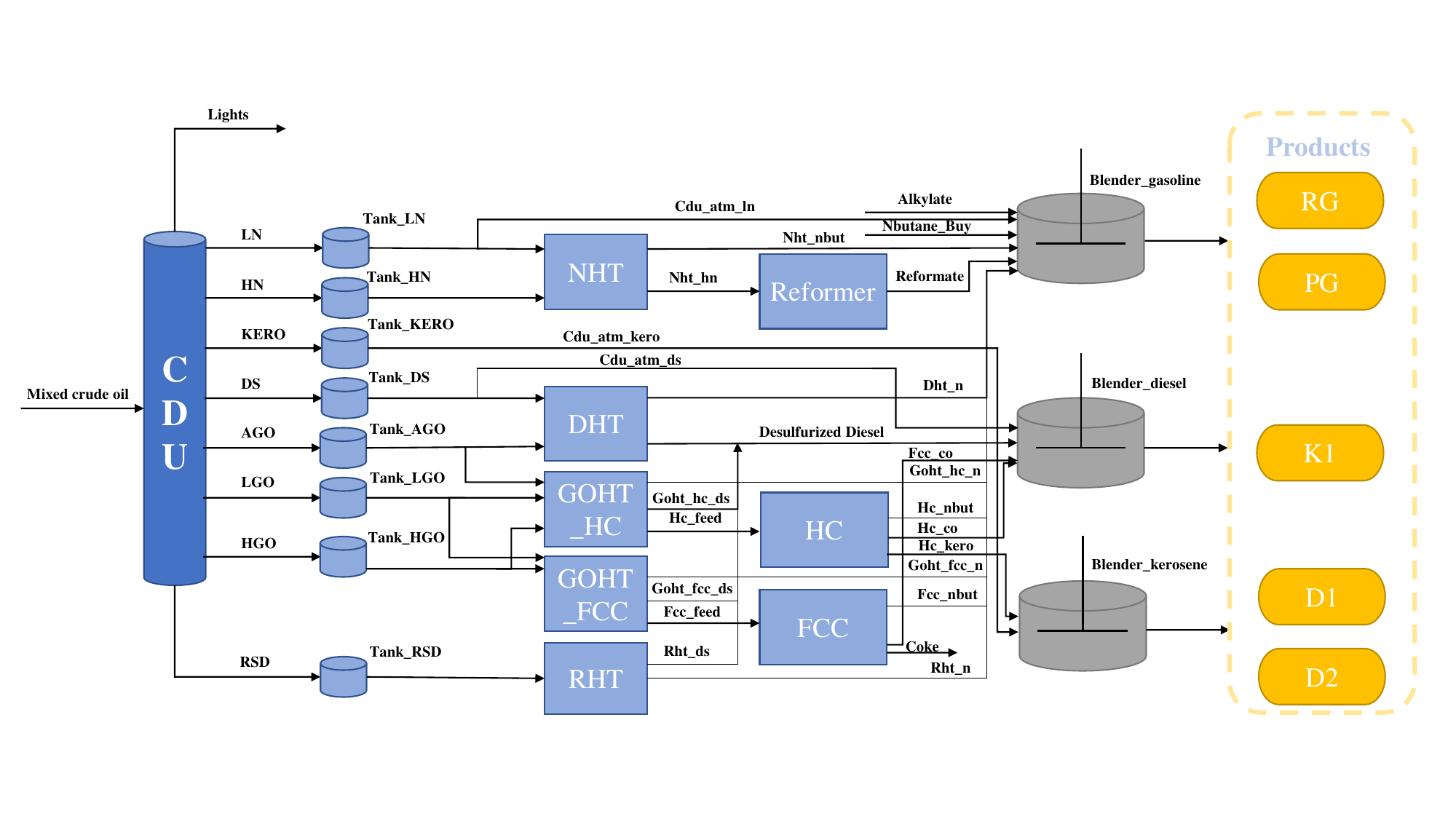}
    \caption{The simplified flowchart of an industrial refinery}
    \label{fig:1}
\end{figure}

Refinery planning is crucial for optimizing production efficiency and ensuring the rational allocation of resources, which encompasses various aspects, including production planning, inventory management, and cost control, and other aspects. A typical refinery planning flow chart can be illustrated in Figure \ref{fig:1}. Initially, the refinery purchases different types of crude oil, represented by the set $C(c\in1,2,...,C)$. These crude oils, each with distinct compositions and properties, are mixed and sent to CDU and other secondary processing units for further processing. After the crude oil undergoes initial distillation in CDU, it is separated into fractions based on their boiling points. These fractions are then further processed by secondary units, such as hydrotreaters, reforming units, hydrocrackers, and fluid catalytic crackers. Some of these secondary units can operate in multiple modes to accommodate complex production demands and enhance operational flexibility. Finally, the output streams are blended in blenders to produce chemical products, including various grades of gasoline, kerosene, and diesel, each of which must meet specific property requirements, as defined by the set $Q(q\in1,2,...Q)$. Additionally, in the context of multi-period refinery planning, inventory management plays a significant role. The set $I$ represents the tanks for intermediate products.

The following parameters are given:
\begin{enumerate}[label=(\arabic*)]
    \item A predefined total planning horizon for refinery operations, denoted by T, and a set of discrete time periods $t=1,2,...T$. Especially, when $T=1$, the refinery planning problem becomes a simple single-period planning problem without inventory tanks.
    \item A set of crude oils, their quality properties and price.
    \item A set of final products, their quality specifications and price given by the market.
    \item A set of storage tanks and their maximum, minimum capacity.
    \item A set of processing units with maximum, minimum capacity, operational modes and manipulating cost.
    \item A set of blenders and their minimum and maximum blending capacity.
    \item The interconnections between all the processing units and storage tanks.
\end{enumerate}

The assumptions of the refinery planning model are provided:
\begin{enumerate}[label=(\arabic*)]
    \item The CDU and other secondary processing units employ the fixed yield method.
    \item The qualities of distillation fractions are computed using linear blending equations on either a volumetric basis or a weight basis.
    \item The output streams of other secondary processing units, except specific gravity and sulfur content which need to be computed, the other qualities are fixed.
    \item The price of final products will vary at different time point while other parameters remain the same.
\end{enumerate}

The following variables of refinery planning model are thus determined:
\begin{enumerate}[label=(\arabic*)]
    \item The amount of each type of crude oil to purchase during time period $t$.
    \item The processed amount, inventory level and qualities of each intermediate products during time period $t$.
    \item The processing amount and operational mode of each secondary processing unit during time period $t$.
    \item The recipe and amount of each blended product during time period $t$.
\end{enumerate}

The objective of refinery planning is to maximize profit, which is derived from the revenue generated by the sales of final products in the market, minus the cost of raw materials, inventory costs for intermediates and products, and the operational costs of the units.

\section{Methodology}\label{Methodology}
\subsection{Framework of optimization method based on DRL and model decomposition}\label{framework}

\begin{figure}
    \centering
    \includegraphics[width=0.9\linewidth]{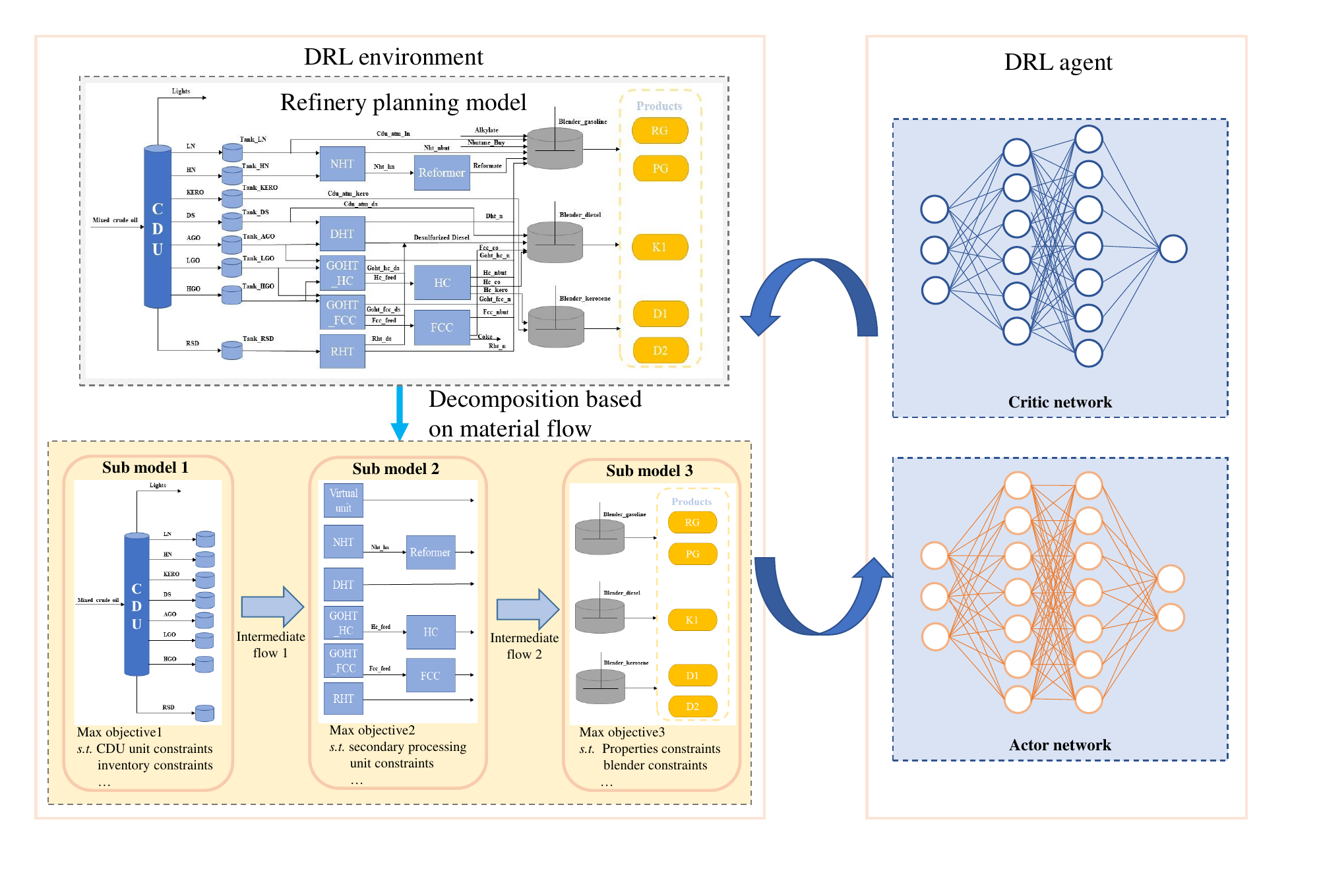}
    \caption{A simplified schematic diagram of the proposed method}
    \label{fig:framework}
\end{figure}
The purpose of this paper is to propose an optimization method for refinery planning based on model decomposition and pricing strategy.This approach aims to enhance the efficiency of solving complex refinery planning models while preserving a certain level of optimality. The refinery planning problem is decomposed into smaller subproblems, each of which features a newly designed objective function called pricing strategy and a subset of the constraints from the overall model. The flowchart of the proposed optimization method, incorporating model decomposition, pricing strategy based on DRL and sequential optimization, is shown in Figure \ref{fig:framework}. The process can be divided into three stages:
\begin{enumerate}
    \item \textbf{Refinery planning model decomposition phase:} The refinery planning model is decomposed into several sub-planning models based on the physical material flow. After defining the appropriate pricing strategy and corresponding constraints, these sub-planning models have the potential to fully replace the overall refinery planning model.
    \item \textbf{Pricing strategy training phase:} The pricing strategy responsible for setting the parameters of the objective function for each sub-planning model is trained as a DRL agent. After defining the necessary elements of the DRL agent, such as states, actions, environments, rewards, network structures and so on, the pricing strategy is trained through historical data and environment.
    \item \textbf{Sequential optimization phase:} Once the pricing strategy has been trained, the appropriate objective function for each sub-planning model can be devised based on the current state. The first sub-planning model then performs optimization through mathematical programming according to the pricing strategy and corresponding constraints, fixes its decision variables, and passes them as inputs to the second sub-planning model. The second sub-planning model will perform similarly, and pass the results to the third sub-planning model, and so on, until all decision variables in the refinery planning process are determined.
\end{enumerate}

The proposed method for solving refinery planning can be summarized in the following steps. First, the refinery planning model is decomposed into several sub-models based on the material flow. Second, the sub-models generated from the decomposition are refined, with particular attention given to their constraint conditions and the design of objective functions. Third, the pricing strategy is developed to interact with the environment and undergo continuous updates. Finally, after training, the pricing strategy can provide optimal parameters $a_t$ for the sub-models' objective functions based on the current state $s_t$.The refinery planning solution is then obtained by solving these sub-models sequentially. The specific implementation of the proposed method is outlined in the following pseudocode:

\begin{algorithm}[H]
  \SetAlgoLined
  \KwData{The refinery planning model, the hyperparameters of pricing strategy}
  \KwResult{All decision variables of refinery planning model}

  Decomposition for refinery planning based on material flow\;

  Designation for objective function to form sub models\;
  
  Initialization the pricing strategy based on DRL with $\pi_\theta$,$v_\phi$ and other hyperparameters \;
  
  \For{$u=1,2,...,U$ where $U$ denotes training total steps}{
  Initialization the state $s_1$\;
  
  \For{$t=1,2,...T$}{
  Sample according to current pricing strategy $a_t=\pi_{\theta_{old}}(s_t)$\;
        
            Interact with environment, $r_t=r(s_t.a_t)$, $s_{t+1}\sim p(s_{t+1}|s_t,a_t)$\;

            Evaluation for value of current policy $v_\phi(s_t)$\;

            Computation of advantage function $A_t(s_t,a_t)$}
            
    Compute loss function of actor and critic network\;
  
  \For{$k=1,2,...K$ where K denotes updating epochs}{
  Updating $\theta$ of actor network\;

  Updating $\phi$ of critic network\;
  }
  $\pi_{\theta_{old}}\xleftarrow{}\pi_\theta$
  }
  \For{$t=1,2,...T$}{
  Get pricing strategy at current moment
  $a_t$=$\pi_\theta(s_t)$\;
  
  Separate pricing parameters for each submodels
  $[\lambda_{1,t},\lambda_{2,t},...\lambda_{n,t}]=a_t$\;
  
  Optimization will begin from submodel $n=1$ \;
  
  \For{$n=1,2,...N$ where N denotes number of sub-planning models}{ 

  Optimize submodel $n$ by mathematical programming to get decision variable $\omega_{n,t}$\;
  The result of submodel $n$ is passed to submodel $n+1$ as input unless $n=N$\;

  Optimize the next submodel $n=n+1$
  }
  All the decision variables at moment $t$ are gained and do $t=t+1$
  }
  All the decision variables of multi-period refinery planning are obtained\;
  
  \caption{Optimization method for refinery planning based on pricing strategy and model decomposition}

\end{algorithm}

\subsection{Rationality of proposed method}\label{Basis}
This approach is applicable to both single-period and multi-period refinery planning. For simplicity, the refinery planning model can be represented as follows:
\begin{equation}
    \begin{aligned}
        min\ p^T\omega \ (\omega\in W) \\
        s.t.\ G(\omega,\beta)\leq 0 \\    
    \end{aligned}
\end{equation}
where the decision variable is $\omega$ and the feasible region $W$ is a mixed-integer set, with $\omega^*$ denoting the optimal solution of the refinery planning model. The parameter $p$ represents cost and as a result, the objective function is linear. However, part of the parameter $p$, specifically the product price, varies over time, while other components, such as crude oil price, operational costs, and inventory costs, remain constant. Additionally, the refinery planning constraints are expressed as $G(\omega,\beta)\leq 0$, with non-linear terms arising from the product properties. The parameter $\beta$ in the constraints is assumed to be constant.

Once all parameters of the refinery planning model are determined, regardless of the scale or complexity of the problem, an optimal solution $\omega^*$ exists for the entire deterministic large-scale planning problem. This solution $\omega^*$ can be used to determine the key variables in refinery planning, including the quantity and properties of each intermediate product, the amount of crude oil to purchase, the operating modes of processing units, the quantities and properties of the final products to be sold, and the inventory levels in the multi-period planning model. This implies that if we can specify these variables, as determined by solving a set of sub-planning problems, the refinery planning model can also yield a global optimal solution.

As shown in Figure \ref{fig:framework}, for sub-model 1, once an objective function is defined, the optimal solution $\omega_1$ can be obtained through mathematical programming with the constraints of the CDU, so that the output streams of CDU is fixed as part of global solution $\omega$. After blending with the inventory tank, the amount and properties of intermediate product 1 are used as input variables, forming the input constraints for sub-model 2. With an appropriate objective function and corresponding constraints, the solution $\omega_2$ for sub-model 2 can be solved, which specifies the amount of intermediate product 2, serving as the input for sub-model 3. As this sequential solving process progresses, the values of the global solution are determined step by step.

In summary, we decompose the refinery planning model into a set of sub-models, where the connections between them are defined by the intermediate products. The constraints of the original planning model are not modified but are distributed across the corresponding sub-models. As a result, the global solution, composed of a set of local solutions $\omega_i$, must be feasible if all sub-models have feasible solution, as these local solutions are within the feasible domain. By designing a suitable objective function for each sub-model, the sub-models can be solved sequentially, leading to the optimal global solution $\omega^*$,which is obtained by combining all the local solutions $\omega^*_i$.

\subsection{Mathematical formulation of proposed method}

In the section \ref{Basis} we give a brief introduction of proposed method which still need to be solved by mathematical programming after decomposing the refinery planning model into sub-planning models. For this study, we take refinery planning model as research object and we will describe the mathematical formulation of proposed method in detail in the following sections, including how to design objective function and the corresponding constraints for sub model.

To accurately and appropriately represent the refinery planning model studied in this paper, several important parameters, indices, sets, and variables will be presented in this section.

The model is based on the discrete-time formulation. Time periods are represented by set $\textbf{T}=\{t\}$, streams are represented by set $\textbf{S}=\{s\}$, crude oils are represented by $\textbf{C}=\{c\}$, final products sold in the market by $\textbf{P}=\{p\}$, blenders by $\textbf{B}=\{b\}$, storage tanks by $\textbf{I}=\{i\}$, all other units by $\textbf{U}=\{u\}$, and the quality properties by $\textbf{Q}=\{q\}$. 

%%, and thus can be classified into subsets $IP_i$ according to the decomposition

Set $\textbf{S}$ include all the intermediate products present in the actual refinery network. Set $\textbf{U}$ not only includes CDU, but also other secondary units, such as hydrotreater, reformer, hydrocracker and other reactors. Inlet streams of units, storage tanks and blenders are respectively represented by sets $\textbf{UI}=\{(u,s)\}$, $\textbf{II}=\{(i,s)\}$, and $\textbf{BI}=\{(b,s)\}$. On the other hand, outlet streams of units, storage tanks and blenders are respectively represented by sets $\textbf{UO}$, $\textbf{IO}$, and $\textbf{BO}$.

\begin{figure}
    \centering
    \includegraphics[width=0.5\linewidth]{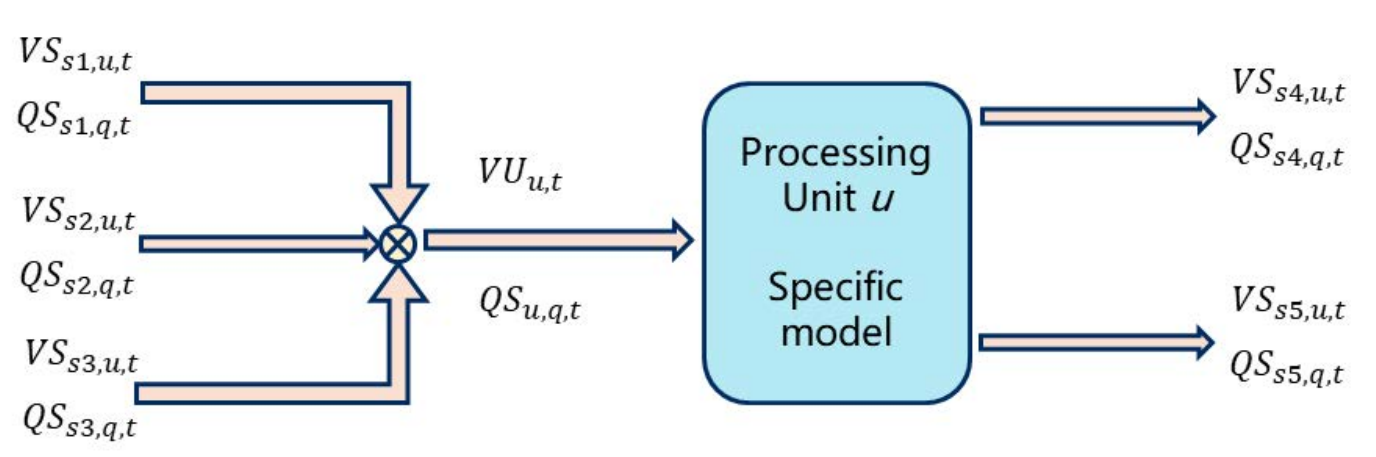}
    \caption{The working mode of the processing unit in the refinery}
    \label{fig:2}
\end{figure}

Besides, the working mode of the processing unit is represented in Figure \ref{fig:2}. When $s:(u,s)\in UI$, the variable $VS_{s,u,t}$ is the inlet volumetric flow rate of stream $s$ feed to unit $u$ during period $t$. The variable $VU_{u,t}$ is the total volumetric flow rate feed to unit $u$ during period $t$. The variable $QS_{s,q,t}$ is the quality property $q$ of $s$ feed to unit $u$ during period $t$. When $s:(u,s)\in UO$, the variable $VS_{s,u,t}$ is the outlet volumetric flow rate of stream $s$ from unit $u$ during period $t$. In the same way, the inlet stream and output stream of the tanks and blenders are represented as $VI_{s,i,t}$ and $VB_{s,b,t}$,respectively.

\subsubsection{Objective function}
For our method, the objective function is not a single function which describe the profit of the refinery as most refinery planning models do, but a well designed function called as pricing strategy for each sub model so that a preferable value of solution $\omega_i$ can be obtained by mathematical programming. The objective function for sub model $i$ is $Obj_i$ which have the universal form as $Eq(2)$:
\begin{equation}
    Obj_{i,t}\ =\ F(\omega_{i,t},\lambda_{i,t})
\end{equation}
where $\omega_{i,t}$ denotes the decision variables of sub model $i$ and $\lambda_{i,t}$ denotes the pricing parameters of $Obj_i$ during period $t$. It is worth noting that there may be many specific forms for $Obj_{i,t}$ as long as the sub model $i$ can maximize $Obj_{i,t}$ when the decision variable equals to $\omega_{i,t}^*$.
And here we introduce some regular forms of $Obj_{i,t}$ for which we will provide an reasonable explanation in \ref{Provement}. For example, it is very common to use the fixed-yield CDU model in the planning problem, 
and we design two simple but useful forms for it:
\begin{equation}
    Obj_{i,t}\ = \sum_{s:(u,s)\in UO}(P_{s,t}VS_{s,u,t}+\sum_{q\in Q}P_{s,q,t}\cdot QS_{s,q,t})\ \ \ \ \ \ \forall t\in T,u\in CDU
\end{equation}
or
\begin{equation}
    Obj_{i,t}\ = \sum_{s:(u,s)\in UO}(P_{s,t}VS_{s,u,t}+\sum_{q\in Q}P_{s,q,t}\cdot QS_{s,q,t})+P_{u,t}VU_{u,t}^2\ \ \ \ \ \ \forall t\in T,u\in CDU
\end{equation}
where $P_{s,t}$ denotes the price of intermediate product $s$ which is defined by ourselves, while $P_{s,q,t}$ and $P_{u,t}$ are, in a sense, the comprehensive price penalty items used to ensure that the quality of the intermediate products meets certain requirements. These terms are essentially equivalent to $\lambda_{1,t}$.

It is also quite simple to design the objective for storage tanks so that it can control its inventory level:
\begin{equation}
    Obj_{i,t} = \tilde{P_{i,t}}\cdot I_{i,t}\ \ \ \ \forall t\in T,i \in I
\end{equation}
\begin{equation}
    \tilde{P_{i,t}} = P_{i,t}^{ref}-\alpha\cdot I_{i,t}\ \ \ \ \forall t\in T,i \in I
\end{equation}
where $P^{ref}_{i,t}$ acts as a reference price parameter which is defined by ourselves, $\alpha$ is a hyper-parameter related to the scale of refinery plant and $I_{i,t}$ denotes the inventory level of intermediate product $i$. $Eq(6)$ means that the price of intermediate product will decrease as the quantity increases.

When the refinery model becomes more complex, it becomes challenging to define an appropriate objective function $Obj_i$ with a simple form that ensures the optimal point exactly equals $\omega_i$ for each sub model $i$ after decomposition. To address this, we introduce a more general approach for constructing the objective function. Specifically, for the last-level sub-models after decomposition, the objective function is defined in the same way as the refinery’s profit, allowing it to optimize itself to the optimum as long as the solutions for the other sub-models are equal to $\omega_i^*$. 

In summary, regardless of the complexity of refinery planning model, there are always sub-models with simpler constraints, such as the CDU, tanks, and secondary processing units. For these units, even if their objective functions are relatively simple, an appropriate parameter setting can ensure that the local optimum exactly matches the global optimum. However, for units with more complex constraints, such as mixers with strict product quality requirements, it is difficult to intuitively construct an objective function framework that satisfies these demands. Therefore, for these complex sub-models, we consider adopting an objective function form similar to that of the overall refinery planning model.

\subsubsection{Processing Unit}
The set $U$ is the collection of processing units in the refinery, including the CDU and other secondary processing units. The operation mode of processing unit can be one or multiple, thus the processing units can be classified into two subset $SU$ and $MU$. The processing units which belong to subset $SU$ have just one operation mode, while those processing units with multiple operation modes belong to subset $MU$. For the real processing unit $ru$ having multiple modes, each operation mode can be represented as a different unit $u$ and these operation modes are represented by set $MMU$. It is worth noting that for multi-operating mode production units, although we allow different operation modes to be run, only a limited number of operating modes are allowed within a production cycle $t$ as shown in $Eq(7)$. An important reason for this is that each transition between operating modes requires both time and cost. 
\begin{equation}
    \sum_{u:(ru,u)\in MUU}bu_{u,t} \leq \ UMM_{ru,t}\ \ \ \ \ \ \  \forall ru\in MU,t\in T
\end{equation}
where the variable $bu_{u,t}$ denotes whether the multi-operating mode processing unit $ru$ operates in a given operating mode $u$ during the production cycle $t$, $UMM_{ru,t}$ is the maximum number of modes in which a real unit $ru$ can operate during period $t$.

For processing units, another important constraint is that the inlet volumetric flow rate must lie between the unit's maximum and minimum processing flow rates during period $t$. 

For the units of $SU$, it is expressed as $Eq(8)$ and $Eq(9)$:
\begin{equation}
    VU_{u,t}=\sum_{s:(u,s)\in UI}VS_{s,u,t}\ \ \ \ \ \ \ \forall u\in SU,t\in T
\end{equation}
\begin{equation}
    V^{min}_{u,t}\leq VU_{u,t}\leq V^{max}_{u,t} \ \ \ \ \ \forall u\in SU,t\in T
\end{equation}
where $VU_{u,t}$ is the flow rate of unit $u$ and $VS_{s,u,t}$ denotes the flow rate of stream $s$ feed to unit $u$ during time period $t$. The processing capacity of unit $u$ is limited by lower bound $V^{min}_{u,t}$ and upper bound $V^{max}_{u,t}$.

While for the units belong to $MU$, the capacity constraint should be modified into $Eq(10)$, $Eq(11)$ and $Eq(12)$:
\begin{equation}
    VU_{ru,t} = \sum_{u:(ru,u)\in MMU}VU_{u,t}=\sum_{s\in S}VS_{s,ru,t}\ \ \ \ \forall ru\in MU,t\in T
\end{equation}
\begin{equation}
    VU_{u,t}=\sum_{s\in S}VS_{s,u,t}\ \ \ \ \forall u:(ru,u)\in MMU,t\in T
\end{equation}
\begin{equation}
    V^{min}_{ru,t}\leq VU_{ru,t}\leq V^{max}_{ru,t}\ \ \ \ \forall ru\in MU,t\in T
\end{equation}

The volumetric flow of an outlet stream $s'$ from unit $u$ is defined as a function of the individual inlet flows $VS_{s,u,t}$, the total feed flow $VU_{u,t}$, the quality of the individual inlet streams $QS_{s,q,t}$ and other operating variables which can be expressed as $Eq(13)$ and $Eq(14)$:
\begin{equation}
\begin{split}
    &VS_{u,s',t}=g_1(VS_{s,u,t},VU_{u,t},QS_{s,q,t},...)\\ 
    &\forall s:(u,s)\in UI,s':(u,s')\in UO,u\in SU, t\in T, q\in Q
\end{split}
\end{equation}
\begin{equation}
\begin{split}
    &VS_{ru,s',t}=g_2(VS_{s,ru,t},VU_{ru,t},QS_{s,q,t},VS_{s,u,t},VU_{u,t},bu_{u,t}...)\ \ \\ &\forall s:(u,s)\in UI,s':(u,s')\in UO,ru\in MU, u:(ru,u)\in MMU,t\in T,q\in Q
\end{split}
\end{equation}

Similarly, the quality $q$ of outlet stream $s'$ from processing unit $u$ is also a function of the individual inlet flows $VS_{s,u,t}$, the total feed flow $VU_{u,t}$, the quality of the individual inlet streams $QS_{s,q,t}$ and other operating variables.
\begin{equation}
\begin{split}
    &QS_{s',q,t}=g_3(VS_{s,u,t},VU_{u,t},QS_{s,q,t},...)\ \ \ \\ 
    &\forall s:(u,s)\in UI,s':(u,s')\in UO,u\in SU, t\in T, q\in Q
\end{split}
\end{equation}
\begin{equation}
\begin{split}
    &QS_{s',q,t}=g_4(VS_{s,ru,t},VU_{ru,t},QS_{s,q,t},VS_{s,u,t},VU_{u,t},bu_{u,t}...)\ \ \ \\ &\forall s:(u,s)\in UI,s':(u,s')\in UO,ru\in MU, u:(ru,u)\in MMU,t\in T
\end{split}
\end{equation}

The actual form of $Eq(13)-(16)$ will depend on the specific processing unit and the model being considered. Some of the models adopted in this paper are presented in the following equations.

\textbf{\emph{3.3.2.1 Crude Distillation Units}}

The CDU model described here is a simple fixed yield model with a single operating mode. Thus the volumetric flow and corresponding quality of outlet stream $s'$ solely depend on the feed composition of CDU. The inlet flow of CDU is composed by various types of crude oils as shown in $Eq(17)$, and $Eq(18)$ computes the size of final cuts which are determined by the fixed yield model based on the assay data of crude oils.
\begin{equation}
    VU_{u,t}=\sum_{c\in C}VC_{c,t}\ \ \ \ \ \forall u\in CDU,t\in T
\end{equation}
\begin{equation}
    VS_{s,u,t}=\sum_{c\in C}Yield_{u,c,s}\cdot VC_{c,t}\ \ \ \ \ \forall s:(u,s)\in UO,u\in CDU,t\in T
\end{equation}

It is assumed that the quality properties of outlet streams are blended linearly, either on a volumetric or on a weight basis, as shown in $Eq(19)$ and $Eq(20)$.

\begin{equation}
    QS_{s',q,t}\cdot VS_{s',u,t}=\sum_{c\in C} QC^{fix}_{q,c,s'}\cdot VC_{c,t}\ \ \ \ \ \forall s':(u,s')\in UO,u\in CDU,q\in SQV
\end{equation}
\begin{equation}
\begin{split}
    &QS_{s',q,t}\cdot QS_{s','sg',t}\cdot VS_{s',u,t}=\sum_{c\in C} QC^{fix}_{q,c,s'}\cdot QC^{fix}_{'sg',c,s'}\cdot VC_{c,t}\ \ \ \ \\ &\forall s':(u,s')\in UO,u\in CDU,q\in SQW
\end{split} 
\end{equation}
where the set $SQV$ denotes the quality properties which are based on volumetric basis, including specific gravity, pour point, aromatic content and so on. And set $SQW$ denotes the quality properties based on weight basis, such as sulfur content.

\textbf{\emph{3.3.2.2 Hydrotreating units}}

The hydrotreating units are considered as units with a single operating mode. The main function of the hydrotreating unit is to reduce the sulfur content in the intermediate product stream, while also improving the quality of the intermediate product to some extent. The unit adopts a fixed yield model, as shown in the following equation.
\begin{equation}
    VS_{s,u,t}=Yield^{fix}_{u,s}\cdot VU_{u,t}\ \ \ \ \ \forall s:(u,s)\in UO, u\in HTU, t\in T
\end{equation}

The quality properties of outlet stream $s$ from hydrotreating units and the sulfur treatment capacity of the hydrotreating units are described by the following equation.
\begin{equation}
    QS_{s,'sg',t}=\alpha\cdot QS_{u,'sg',t}\ \ \ \ \ \forall s:(u,s)\in UO, u\in HTU,t\in T
\end{equation}
\begin{equation}
    QS_{s,q,t}=QS^{fix}_{s,q,t}\ \ \ \ \ \forall s:(u,s)\in UO, u\in HTU,t\in T, q\in SQ \bigwedge q\notin \{'sg','sul'\}
\end{equation}
\begin{equation}
    WS_{s,u,t}=QS_{s,'sg',t}\cdot VS_{s,u,t}\ \ \ \ \forall s:(u,s)\in UI,u\in HTU, t\in T
\end{equation}
\begin{equation}
    WS_{s',u,t}=QS_{s','sg',t}\cdot VS_{s',u,t}\ \ \ \ \forall s':(u,s')\in UO,u\in HTU, t\in T
\end{equation}
\begin{equation}
    SulS_{s,u,t}=QS_{s,'sul',t}\cdot WS_{s,u,t}\ \ \ \ \forall s:(u,s)\in UI,u\in HTU, t\in T
\end{equation}
\begin{equation}
    SulS_{s',u,t}=QS_{s','sul',t}\cdot WS_{s',u,t}\ \ \ \ \forall s':(u,s')\in UO,u\in HTU, t\in T
\end{equation}
\begin{equation}
    \sum_{s:(u,s)\in UI}SulS_{s,u,t}=\sum_{s':(u,s')\in UO}SulS_{u,s',t}+RSUL_{u,t}\ \ \ \ \forall u\in HTU,t\in T
\end{equation}
\begin{equation}
    RSUL_{u,t}\leq RSUL_{u,t}^{max}\ \ \ \ \forall u\in HTU, t\in T
\end{equation}
\begin{equation}
\begin{split}
    &(1-HTU_u^{max})QS_{s,'sul',t}\leq QS_{s','sul',t}\leq (1-HTU_u^{min})QS_{s,'sul',t}\ \ \ \\ &\forall u\in HTU,s:(u,s)\in UI,s':(u,s')\in UO,t\in T
\end{split}
\end{equation}
where $\alpha$ is a hyper-parameter and is set to 0.98. The $QS_{u,q,t}$ is quality property $q$ of the total inlet flow of unit $u$ during period $t$. Parameter $WS_{s,u,t}$ denotes the weight of inlet stream $s$ which flow into unit $u$. $Eq(27)-(28)$ represent the sulfur content of stream $s$. It is a reasonable assumption that the sulfur removal capacity of the hydrotreating unit has an upper limit which is shown as $Eq(30)-(31)$.

\textbf{\emph{3.3.2.3 Other secondary processing unit}}

In our refinery planning model, the other secondary processing units belong to units with multiple operating modes. For these units, the flow into the real processing unit $ru\in MU$ is equal to the sum of the flows into each operating mode of the multi-mode processing unit $u:(ru,u)\in MMU$, which is shown as $Eq(10)$. The flow rate and quality properties of outlet stream from secondary processing units are given by the following equations.
\begin{equation}
    VS_{s,t}=Yield^{fix}_{u,s}VU_{u,t}\ \ \ \ \forall s:(u,s)\in UO, u:(ru,u)\in MMU, t\in T
\end{equation}
\begin{equation}
    QS_{s,q,t}=QS_{s,q,t}^{fix} \ \ \ \ \forall s:(u,s)\in UO, u:(ru,u)\in MMU, t\in T,q\in SQV
\end{equation}
\begin{equation}
\begin{split}
    &QS_{s',q,t}=SulR_{u,t}^{fix}\cdot QS_{s,q,t} \ \ \ \\ \forall s':(u,s')\in UO,&s:(u,s)\in UI, u:(ru,u)\in MMU, t\in T,q\in SQW
\end{split}
\end{equation}

\subsubsection{Storage Tanks}
In our refinery planning model, there are several storage tanks for intermediate products from CDU as shown in Figure \ref{fig:1}. And we assume that there is no intermediate product in the beginning, i.e. $I^{ini}_i=0$. The equations of material balance for storage tanks are given by $Eq(35)-(36)$. :
\begin{equation}
    \sum_{s:(i,s)\in II}VI_{s,i,t}+I^{ini}_{i}=I_{i,t}+\sum_{s':(i,s')\in IO}VI_{s',i,t}\ \ \ \ \forall t=1, i\in I
\end{equation}
\begin{equation}
    \sum_{s:(i,s)\in II}VI_{s,i,t}+I_{i,t-1}=I_{i,t}+\sum_{s':(i,s')\in IO}VI_{s',i,t}\ \ \ \ \forall t>1, i\in I
\end{equation}

And the inventory level of intermediate product must be between the minimum and maximum limits:
\begin{equation}
    IL_i^{min}\leq I_{i,t} \leq IL_i^{max} \ \ \ \ \forall i\in I
\end{equation}

Besides, for all the possible quality properties of the inlet and outlet streams, the tanks mixes them according to the linear mixing law.
\begin{equation}
\begin{split}
    \sum_{s:(i,s)\in II}VI_{s,i,t}\cdot &QS_{s,q,t}+I^{ini}_{i}\cdot QI^{ini}_{i,q}=\\I_{i,t}\cdot QI_{i,q,t}+&\sum_{s':(i,s')\in IO}VI_{s',i,t}\cdot QS_{s',q,t}\ \ \ \\ &\forall t=1, i\in I, q\in SQV
\end{split}
\end{equation}
\begin{equation}
\begin{split}
    \sum_{s:(i,s)\in II}VI_{s,i,t}\cdot &QS_{s,q,t}+I_{i,t-1}\cdot QI_{i,q,t-1}=\\I_{i,t}\cdot QI_{i,q,t}+&\sum_{s':(i,s')\in IO}VI_{i,s',t}\cdot QS_{s',q,t}\ \ \ \\ &\forall t>1, i\in I, q\in SQV
\end{split}
\end{equation}
\begin{equation}
\begin{split}
    \sum_{s:(i,s)\in II}VI_{s,i,t}\cdot QS_{s,'sg',t}\cdot QS_{s,q,t}&+I^{ini}_{i}\cdot QI^{ini}_{i,'sg'}\cdot QI^{ini}_{i,q}=\\I_{i,t}\cdot QI_{i,'sg',t}\cdot QI_{i,q,t}+\sum_{s':(i,s')\in IO}&VI_{s',i,t}\cdot QS_{s','sg',t}\cdot QS_{s',q,t}\ \ \ \\ &\forall t=1, i\in I, q\in SQW
\end{split}
\end{equation}
\begin{equation}
\begin{split}
    \sum_{s:(i,s)\in II}VI_{s,i,t}\cdot QS_{s,'sg',t}\cdot &QS_{s,q,t}+I_{i,t-1}\cdot QI_{i,'sg',t-1}\cdot QI_{i,q,t-1}=\\I_{i,t}\cdot QI_{i,'sg',t}\cdot QI_{i,q,t}+&\sum_{s':(i,s')\in IO}VI_{i,s',t}\cdot QS_{s','sg',t}\cdot QS_{s',q,t}\ \ \\ \ &\forall t>1, i\in I, q\in SQW
\end{split}
\end{equation}
\subsubsection{Blenders}
For the blenders, in addition to the material balance and linear mixing law, it is important to note that although a blender can produce multiple products, only one product can be produced within a given production cycle $t$. Moreover, this product must meet the strict quality constraints required by the external market.
\begin{equation}
    VB_{b,t}=\sum_{s:(b,s)\in BI}VS_{s,b,t}
\end{equation}
\begin{equation}
    VB_{b,t}^{min}\leq VB_{b,t}\leq VB_{b,t}^{max}\ \ \ \ \forall b\in B,t\in T 
\end{equation}
\begin{equation}
    \sum_{p:(b,p)\in BO}bb_{p,t}=1\ \ \ \ \forall b\in B,t\in T
\end{equation}
\begin{equation}
    \sum_{p:(b,p)\in BO} VP_{p.t}\cdot bb_{p.t}=VB_{b,t}\ \ \ \ \forall b\in B,t\in T
\end{equation}
\begin{equation}
    \sum_{p:(b,p)\in BO}VP_{p,t}\cdot QP_{p,q,t}=\sum_{s:(b,s)\in BI}VS_{s,b,t}\cdot QS_{s,q,t}\ \ \ \ \forall b\in B,q\in SQV
\end{equation}
\begin{equation}
\begin{split}
    \sum_{p:(b,p)\in BO}VP_{p,t}\cdot &QP_{p,'sg',t}\cdot QP_{p,q,t}=\sum_{s:(b,s)\in BI}VS_{s,b,t}\cdot QS_{s,'sg',t}\cdot QS_{s,q,t}\ \ \\ \ &\forall t\in T,b\in B,q\in SQW
\end{split}
\end{equation}
\begin{equation}
    QP^{L}_{p,q,t}\leq QP_{p,q,t} \leq QP^{U}_{p,q,t}\ \ \ \forall t\in T,q\in Q,p\in P
\end{equation}

\subsection{Pricing strategy based on DRL}

In the previous section, we provided a detailed description of the mathematical model and the corresponding physical meanings of the proposed method. Once the model decomposition approach and the objective function form for the sub-models are defined, the relevant parameters of the sub-models can be determined for deterministic models. Specifically, this study focuses on the refinery planning problem, where external product prices fluctuate over time. When external prices change, the optimal solution to the planning problem must also adjust accordingly, leading to changes in purchased crude oils, operating conditions, and storage levels. In such cases, the objective function parameters of the sub-planning models must also be updated. However, directly capturing the mapping relationship of such a model, with its complex topology, nonlinearity, and numerous constraints, is highly challenging. Therefore, this study adopts a DRL-based strategy, referred to as the pricing strategy, to capture the relationship between external price fluctuations and the parameters of the sub-planning models’ objective functions. The pricing strategy enables refinery to make informed decisions while enhancing the speed and effectiveness of solving the planning problem.

DRL is a learning paradigm aimed at maximizing cumulative rewards. The pricing strategy continuously interacts with the environment $E$, receiving a reward signal to adjust its behavior accordingly. The Markov Decision Process (MDP) is particularly well-suited for modeling RL tasks, where $\mathscr{S}$ denotes the state space and $\mathscr{A}$ stands for the action space. At each step, the pricing strategy will takes an action $a_t$ based on the current state $s_t$ and policy $\pi_\theta(a_t|s_t)$ to interact with the environment. It then receives the corresponding reward $r_t(a_t,s_t)$ and updates the state $s_{t+1}$ accordingly.

Specifically, the episode reward $R_t$ is calculated by the discounted sum of immediate rewards $r_t$:
\begin{equation}
    R_t=\sum_{t=1}^T\gamma^tr_t=\sum_{t=1}^T\gamma^tr(s_t,a_t)
\end{equation}

The RL objective is defined as:
\begin{equation}
    J(\pi_\theta)=max_{\pi_\theta} \mathbb{E}_{s\sim \rho^\pi,a\sim\pi_\theta}[R_t|s_t,a_t]
\end{equation}

To evaluate the outcome of a given state or state-action pair, a value function is introduced in DRL. Typically, the state-action value function denoted as the expected value of episode reward $R_t$ under the current policy $\pi$ is widely used:
\begin{equation}
    Q^\pi(s_t,a_t)=\mathbb{E}_\pi [R_t|s_t,a_t]
\end{equation}

For this paper, the objective of pricing strategy is to maximize the profit earned by the refinery. In the DRL training phase, we assume that proper decomposition has been performed and the forms of the objective functions $Obj_i$ have been defined. The market product prices vary over time, and the inventory levels and quality properties of the tanks may change based on decisions made earlier. Additionally, we assume that the uncertainty regarding future product prices is known, and there are no constraints limiting the sale of products. Therefore, the state of pricing strategy is formulated as $\mathscr{S}\triangleq[P_{p,t},\hat{P}_{p,t+1},I_{t-1},QS_{i,q,t}]$, where $P_{p,t}$ is the current product price, $\hat{P}_{p,t}$ is the estimate of the product price at the next moment, $I_{t-1}$ represents the previous inventory level, and $QS_{i,q,t}$ represents the quality properties of the intermediate products. While the action of pricing strategy $\mathscr{A}\triangleq [\lambda_{1,t},\lambda_{2,t},...\lambda_{N,t}]$ is defined as the input, specifically the pricing parameters of objective function $Obj_i$ of each sub model.

In the framework of this study, the reward function represents the profit earned by the refinery using the proposed method. The reward is not obtained from a single comprehensive mathematical programming calculation; instead, it is determined incrementally as each sub-model is solved sequentially. The values of the refinery planning variables are progressively determined, and once all variables are finalized, the total reward for the refinery planning is obtained. However, since infeasible solutions may exist, a penalty term is added to the reward function to discourage solutions of poor quality.
\begin{equation}
    r_t = R(\omega_{1,t},...\omega_{n,t})+Penalty_t=Profit_t(\omega_{1,t},...\omega_{n,t})+Penalty_t
\end{equation}

\begin{equation}
    Penalty_t =
    \begin{cases}
    -M, & \text{if} \ \ (\omega_{1,t},...\omega_{n,t})\ \text{is infeasible},\\
    0, & \text{if }\ \ (\omega_{1,t},...\omega_{n,t})\ \text{is feasible}
    \end{cases}
\end{equation}
where $-M$ is a quite large number to punish the infeasible solution $\omega_{1,t},...\omega_{n,t}$, otherwise $Penalty_t$ equals to 0.

Proximal Policy Optimization (PPO) is a classical and widely applicable DRL algorithm \cite{schulman2017proximal}. It updates the policy function by collecting multiple trajectories and evaluates state values using a value function. The clipping technique is extensively employed in PPO to ensure that the difference between the new and old policies remains within a specified range during the policy update, thereby ensuring both good performance and stability.

\begin{figure}
    \centering
    \includegraphics[width=1\linewidth]{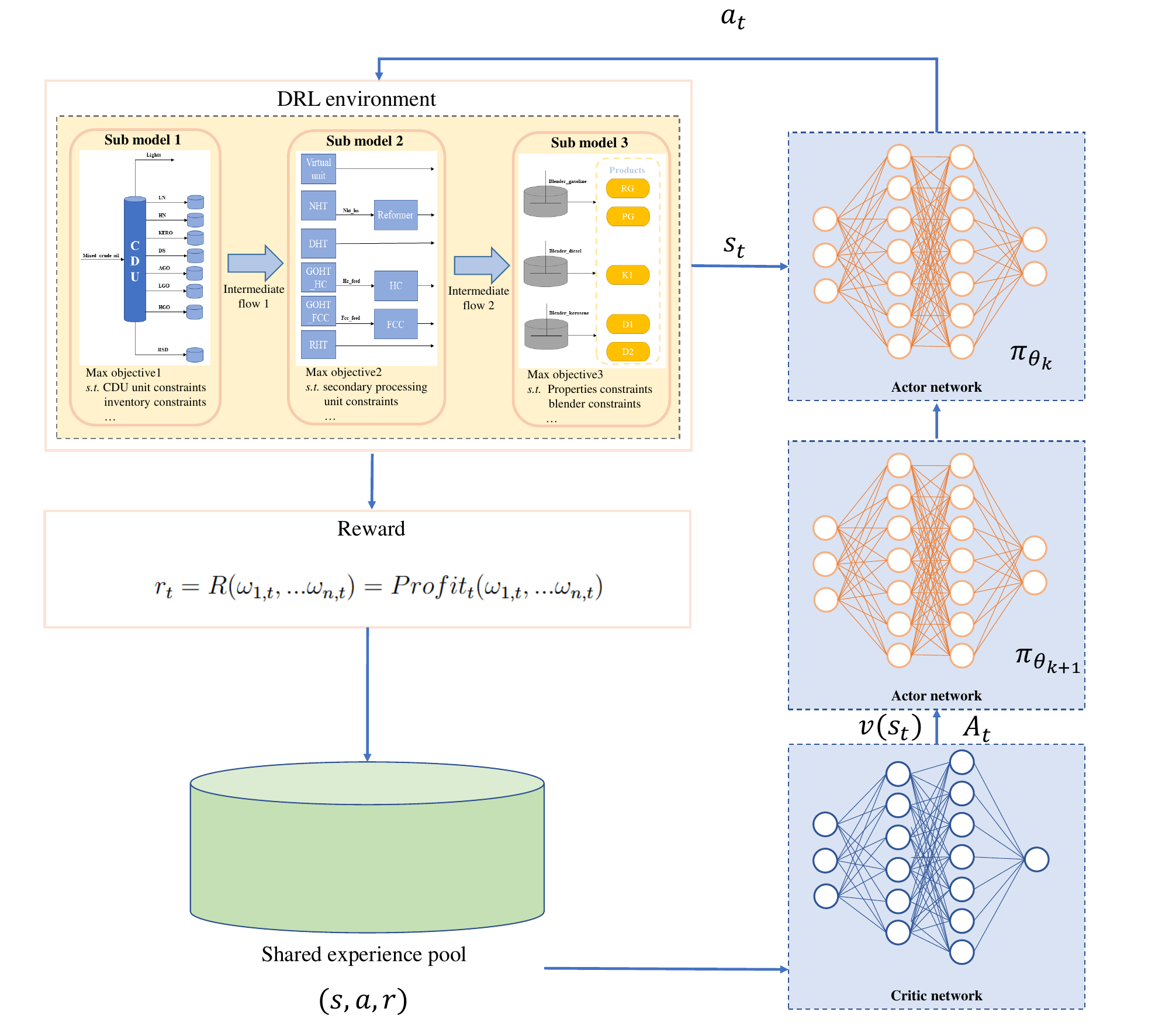}
    \caption{The structure and one update step DRL-based pricing strategy }
    \label{fig:3}
\end{figure}

In this study, we utilize the PPO-Clip algorithm. The process of the pricing strategy completing one learning episode includes the following steps as shown in Figure \ref{fig:3}:

\begin{enumerate}[label=(\arabic*)]
    \item The actor network interacts with the environment $E$ based on the policy $\pi_{\theta_{old}}$ updated in the previous iteration, obtaining the current state-action pair and the corresponding reward. This information $(s,a,r)$ is then stored in a shared experience pool for subsequent agent optimization.
    \item The critic network estimates the value of the current policy, calculates the future discounted rewards, and computes the advantage function which compares the difference between the expected reward of taking a specific action and the average expected reward of following the current policy in the current state.
\begin{equation}
    v_\phi(s_t)\approx E[r_t+\gamma r_{t+1}+\gamma^2r_{t+2}+...+\gamma^Tr_{t+T}]
\end{equation}
\begin{equation}
    A_t(s_t,a_t)=\sum_{i=t}^T\gamma^{i-t}r_i-v_\phi(s_t)
\end{equation}
    \item Then the actor network is updated, and the function for updating policies is shown as:
\begin{equation}
    \theta_{k+1}=argmax_\theta \mathbb{E}_{s,a\sim\pi_{\theta_{k}}}[L(s,a,\theta_k,\theta)]
\end{equation}
where the term $L$ is defined as:
\begin{equation}
    L(s,a,\theta_k,\theta)=min(\frac{\pi_\theta(a|s)}{\pi_{\theta_k}(a|s)}A^{\pi_{\theta_k}}(s,a),clip(\frac{\pi_\theta(a|s)}{\pi_{\theta_k}(a|s)},1-\epsilon,1+\epsilon)A^{\pi_{\theta_k}}(s,a))
\end{equation}
where $\theta_k$ represent the policy parameters before updating at iteration $k$, $\epsilon$ is a hyperparameter used to prevent the policy parameters update from being too large. $\pi_\theta(a|s)$ and $\pi_{\theta_k}(a|s)$ are the importance sampling ratio.
    \item The critic network also needs to be updated. Its update mechanism is based on the loss function and the learning method of the temporal difference (TD) residual. The loss function is represented as follows:
\begin{equation}
    L(\phi)=-(r_t+\gamma v_\phi(s_{t+1})-v_\phi (s_t))^2
\end{equation}
    \item Finally, the Adam optimization algorithm is used to optimize the accumulated loss. The weights of the neural network are iteratively updated based on the training data, allowing for the design of independent adaptive learning rates for different parameters.
\end{enumerate}

\section{Case study}\label{Case study}
This section presents three industrial cases to access the effectiveness of the proposed method. All experiments were conducted using Python 3.8 on a computer equipped with an Intel Core i5-13400F CPU @ 2.50GHz and 16.0 GB of RAM. The DRL-based pricing strategy was trained using PyTorch-lightning 2.4.0, and the optimization solver was Gurobi 11.0.2.

\subsection{Single-period MINLP refinery planning case}
The refinery's production planning process involves importing five types of crude oil, which are fully mixed and then sent to the CDU for processing, where various intermediate products are produced according to the boiling points of crude oils. These intermediate products are further processed in secondary processing units to improve certain quality properties. Finally, these outlet streams are blended in the blenders to produce the final products sold in the market, which include two types of gasoline, one type of kerosene, and two types of diesel. The market imposes strict quality requirements for these products, which are evaluated based on specific gravity, research octane number, motor octane number, Reid vapor pressure, aromatic content, sulfur content, cold injection nozzle, and pour point, as shown in Table \ref{tab:1}. 

For the hyperparameters of pricing strategy based on DRL, we fix the number of hidden layers in critic and actor network at 128 and Adam Optimizer was utilized. The learning rates for the actor network and critic network were set to $7*10^{-5}$ and $2*10^{-4}$, respectively. Other parameters are as follows: a discount factor of 0.9 for the reward, a batch size of 10, 2000 for training episodes, and a clipping rate $\epsilon$ of 0.2. 

\begin{table}[ht]
    \centering
    \caption{Product Quality Specifications}
    \begin{tabular}{ccccccccc}
    \textbf{Product} & \textbf{Sg} & \textbf{RON} & \textbf{MON} & \textbf{RVP} & \textbf{Arom} & \textbf{Sul} & \textbf{CIN} & \textbf{Pour} \\
    \hline
    & & & & \textbf{Minimum}& & & & \\
    RG & 0.73 & 88.5 & 78.5 & 0 & 0 & 0 & NA & NA \\
PG & 0.73 & 92.5 & 82.5 & 0 & 0 & 0 & NA & NA \\
K1 & 0.75 & NA & NA & NA & NA & 0 & NA & 0 \\
D1 & 0.81 & NA & NA & NA & NA & 0 & 40 & 0 \\
D2 & 0.81 & NA & NA & NA & NA & 0 & 40 & 0 \\
\hline
 & & & & \textbf{Maximum} & & & & \\
RG & 0.81 & 150 & 150 & 15 & 60 & 0.001 & NA & NA \\
PG & 0.81 & 150 & 150 & 15 & 60 & 0.001 & NA & NA \\
K1 & 0.85 & NA & NA & NA & NA & 0.3 & NA & 407 \\
D1 & 0.87 & NA & NA & NA & NA & 0.0015 & 100 & 470 \\
D2 & 0.87 & NA & NA & NA & NA & 0.0015 & 100 & 456 \\
    \end{tabular}
    
    \label{tab:1}
\end{table}

\begin{figure}
    \centering
    \includegraphics[width=0.75\linewidth]{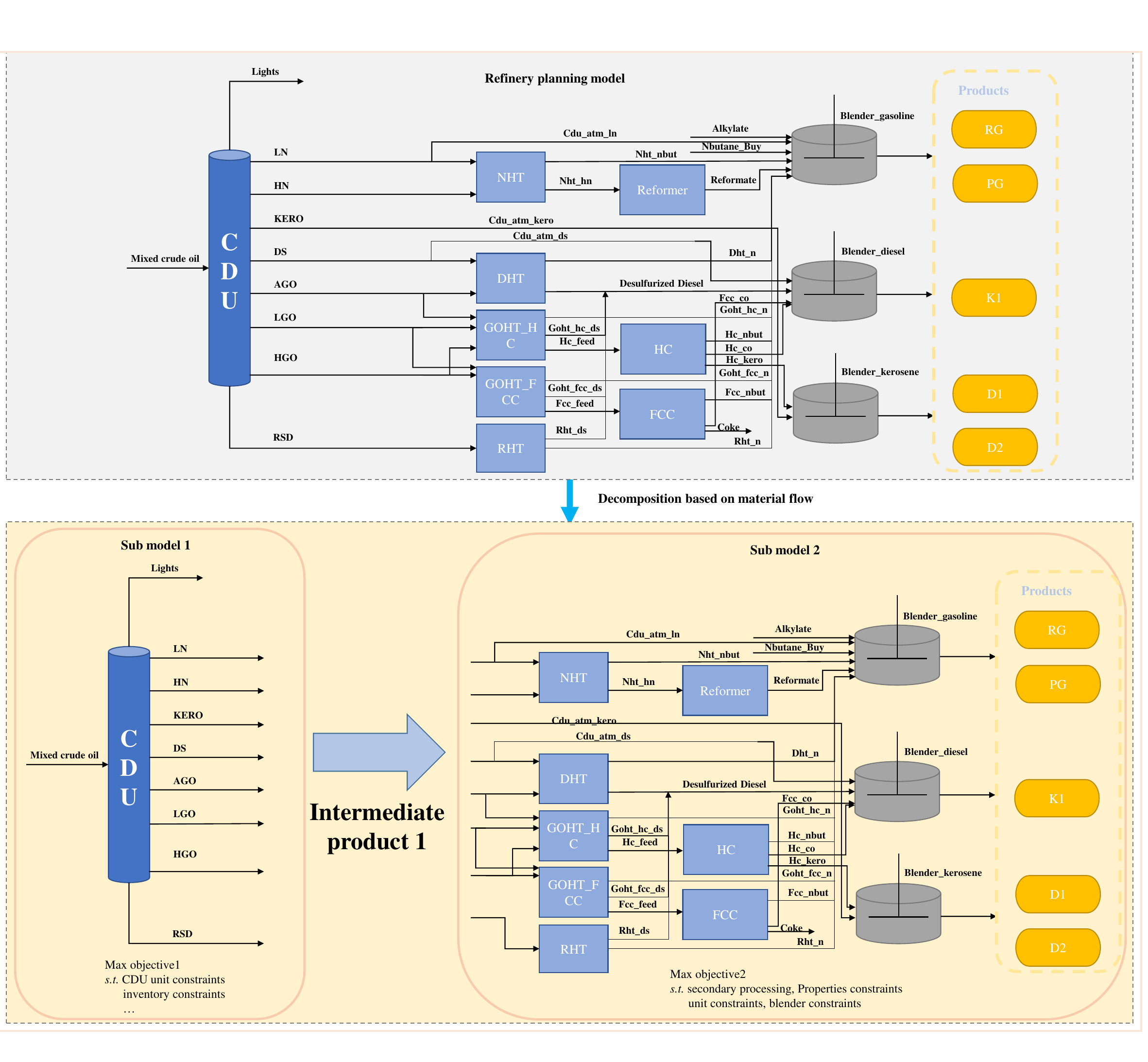}
    \caption{The decomposition method for Case 1}
    \label{fig:decomposition for case 1}
\end{figure}

To effectively apply the proposed method to this single-period MINLP model, whose parameter are referenced from castillo's model \cite{castillo2017global}, the objective function for the sub-planning model 1 and 2 are designed according to the refinery planning network and its decomposition method shown in Figure \ref{fig:decomposition for case 1} as follows:
\begin{equation}
    Obj_1=F(\omega_i,\lambda_{i})=\sum_{s:(u,s)\in UO}(P_s\cdot VS_s + \sum_{q\in QS} P_{s,q}\cdot QS_{s,q} ) \ \ \ \forall u\in CDU
\end{equation}
where $P_s$ and $P_{s,q}$ are the specific implementation of $\lambda$, they can be interpreted as the prices of the intermediate products and the comprehensive penalty factors for product quality, respectively. In this way, the sub model can adjust the crude oil feed based on the intermediate price provided by pricing strategy, ensuring the yield and quality properties of the intermediate products.
\begin{equation}
    Obj_2=\sum_{p\in P}P_{p}\cdot VP_{p}\ -\ \sum_{c\in C}P_{c}\cdot VC_{c} -
    \ \sum_{u\in U}P_{u}\cdot VU_{u}\ -\sum_{m\in M}P_m\cdot VS_m
\end{equation}
\begin{equation}
    Profit=\sum_{p\in P}P_{p}\cdot VP_{p}\ -\ \sum_{c\in C}P_{c}\cdot VC_{c} -
    \ \sum_{u\in U}P_{u}\cdot VU_{u}\ -\sum_{m\in M}P_m\cdot VS_m
\end{equation}
where $M$ represents the raw materials other than crude oil used in the refinery planning, specifically referring to $Alkylate$ and $n-butane\_buy$. Although the objective function of the sub model 2 is similar to that of the refinery planning model, the decomposition approach allows certain key decision variables, especially the flow rate and quality of the CDU product stream, to be determined in advance. This significantly improves the computational efficiency of the solution process.

The important economic data is shown in Table \ref{tab:2}. Since our goal is to study the refinery's decision strategy under external price fluctuations, which is a very common scenario in refinery planning, we assume that external prices are uniformly distributed within a certain range, centered around the given data.
\begin{equation}
    P \sim U(P_{\text{ref}} - \Delta P, P_{\text{ref}} + \Delta P)
\end{equation}

\begin{table}[htbp]
\centering
\caption{Economic Data}
\begin{tabular}{|l|c|l|c|}
\hline
\textbf{Raw Material} & \textbf{$CS_s$ (\$/bbl)} & \textbf{Product} & \textbf{$PS_s$ (\$/bbl)} \\
\hline
CO1 & 40 & RG & 95 \\
CO2 & 41 & PG & 108 \\
CO3 & 42 & K1 & 97 \\
CO4 & 38 & D1 & 100 \\
CO5 & 36 & D2 & 120 \\
alkylate & 129 & $lc_coke$ & 0 \\
n-butane & 32 & LPG & 0 \\
\hline
\end{tabular}

\label{tab:2}
\end{table}

In this case, the state space, action space, and reward are essentially the refinery's external market prices, the pricing parameters of the refinery's sub models, and the overall profit of the refinery, respectively, i.e.$\mathscr{S}\triangleq[P_{p}]$,$\mathscr{A}\triangleq[P_{s},P_{s,q}]$,$R=Profit$

\begin{figure}
    \centering
    \includegraphics[width=0.8\linewidth]{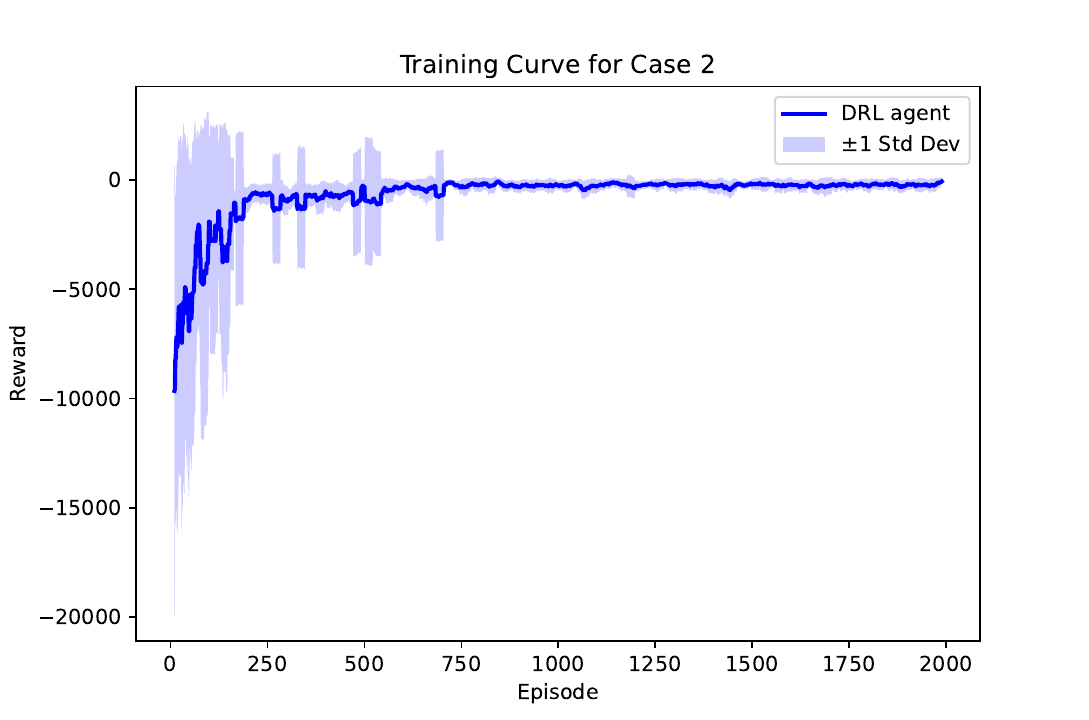}
    \caption{The training curve for pricing strategy in Case 1}
    \label{fig:training curve case1}
\end{figure}

Figure \ref{fig:training curve case1} visualizes the training curve of the proposed algorithm.  Despite having 2000 training episodes with a batch size of 3, the total training time is approximately 15 minutes. It is evident that the algorithm rapidly converges after about 750 episodes and maintains stability thereafter.

\begin{figure}
    \centering
    \includegraphics[width=0.75\linewidth]{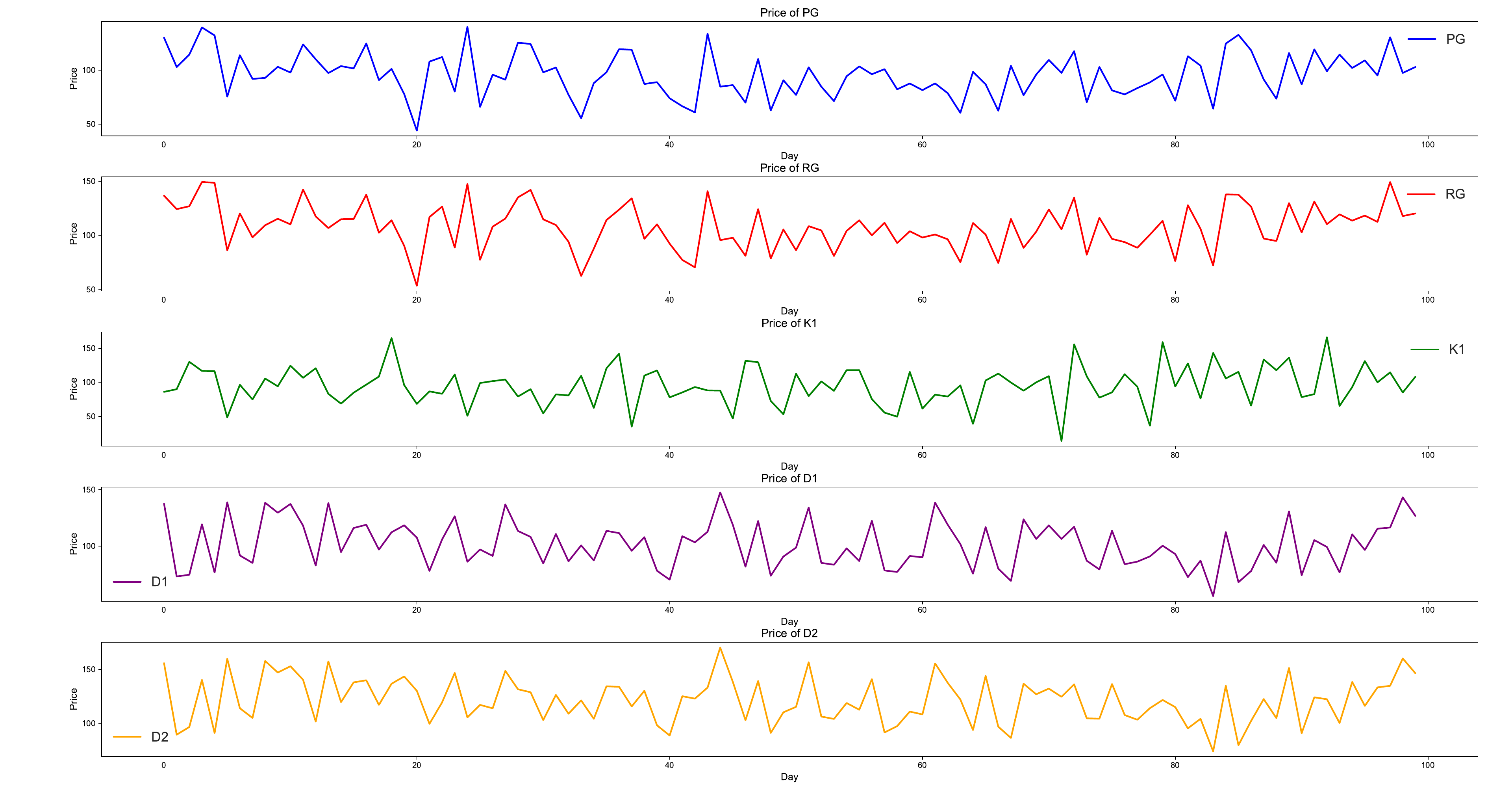}
    \caption{Historical price of PG, RG, K1, D1 and D2}
    \label{fig:price_case1}
\end{figure}

To compare the performance of the method we proposed with other classical methods, we conducted the experiment to compare the refinery planning profit earned by our method, a deterministic model based on mathematical programming as well as a real-time optimization method based on mathematical programming. In the deterministic model, the price for each product is set to the average level. In contrast, the real-time prices at each time step are available for the other methods, allowing the real-time optimization model to always obtain the optimal solution each day. As illustrated in Figure \ref{fig:price_case1}, the daily price of refinery products fluctuates over the course of the 100-day episode. For instance, the prices of gasoline and diesel drop to a low point around the 60th and 85th day, and the price of diesel exhibits multiple peaks on the 10th and 50th day. The volatility of these petroleum product market prices highlights the need for a responsive dynamic planning solution to mitigate the risks arising from improper refinery planning strategies due to external market price fluctuations and challenges the pricing strategy's ability to effectively respond to price fluctuations.

\begin{figure}
    \centering
    \includegraphics[width=0.75\linewidth]{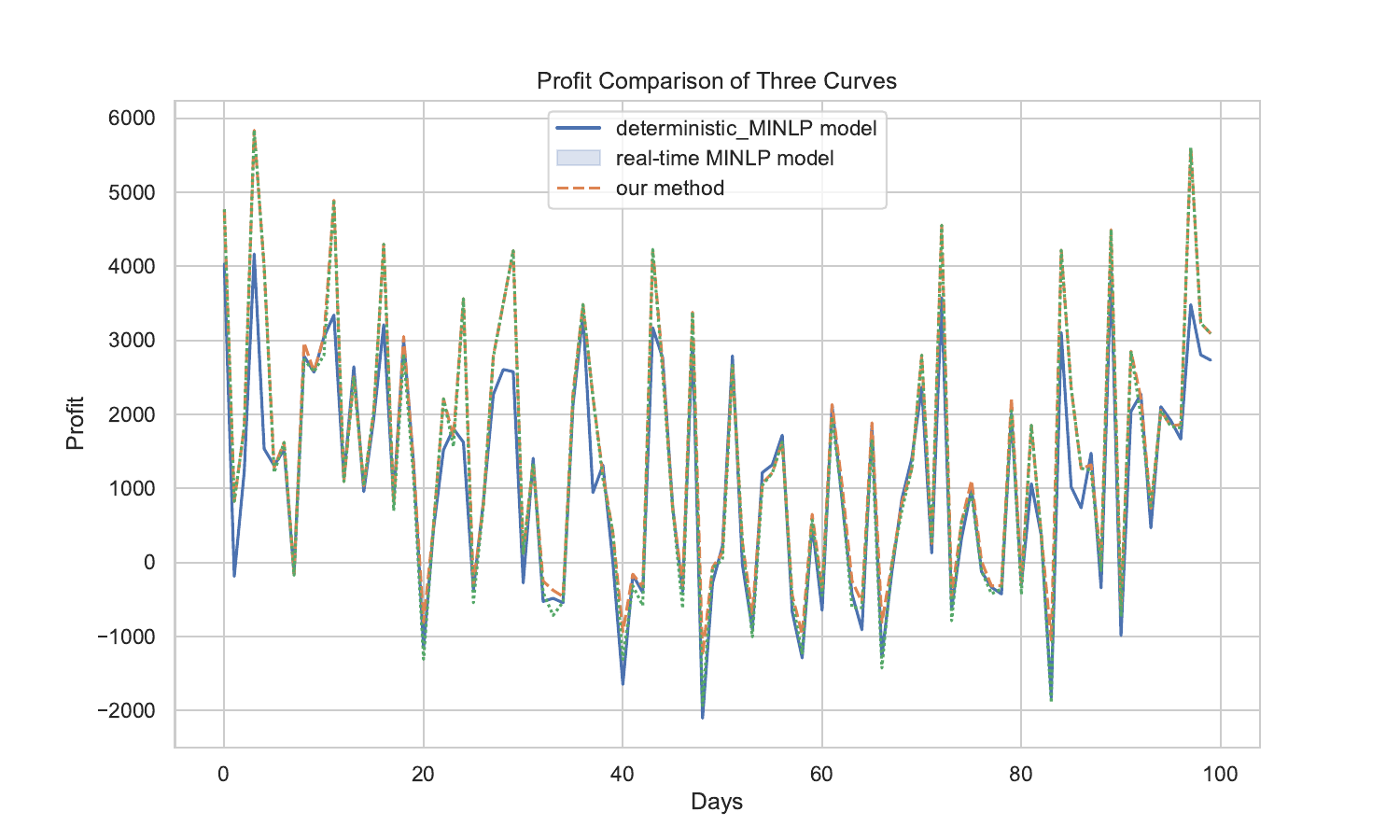}
    \caption{The difference of daily profit of refinery planning based on the three methods}
    \label{fig:profit_case1}
\end{figure}

\begin{figure}
    \centering
    \includegraphics[width=1\linewidth]{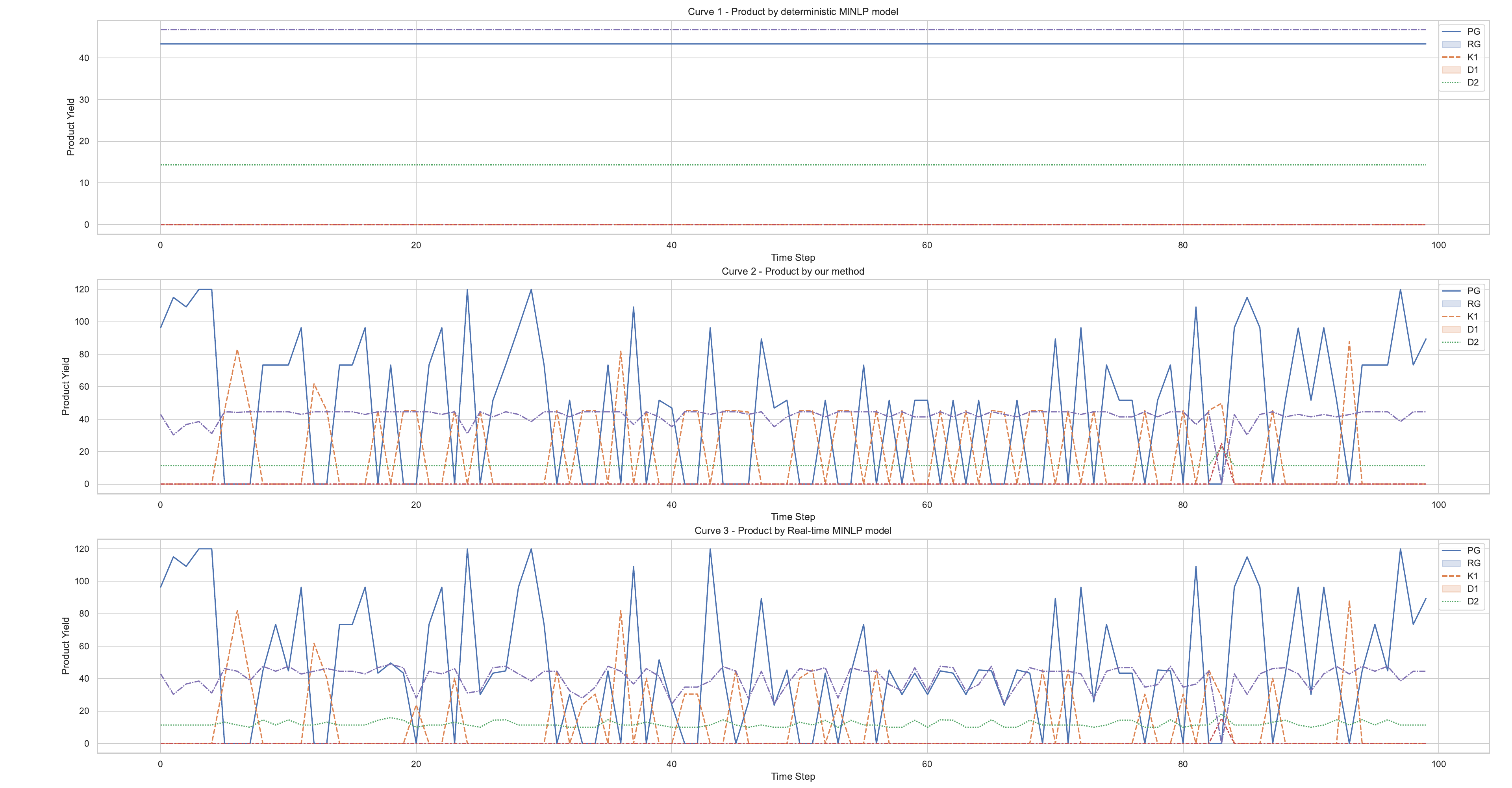}
    \caption{The refinery planning strategy based on the three methods}
    \label{fig:strategy_case1}
\end{figure}

As shown in Figure \ref{fig:strategy_case1}, the three curves represent the refinery planning strategies based on a deterministic model, a real-time mathematical programming model, and the proposed method, respectively, with daily profits over a 100-day period as shown in Figure \ref{fig:profit_case1}. Since the deterministic MINLP model does not take fluctuations into account, planning results remain constant throughout the time period. The real-time MINLP model, however, can always obtain the optimal solution as it performs real-time optimization based on the daily price levels. For our method, we observe that in most cases, its performance is equivalent to that of the real-time mathematical programming model, with only a few instances resulting in suboptimal solutions.

The performance comparison of the three algorithms, shown in Table \ref{Performance Comparision of three method}, reveals that although the constraints and variable scales of the proposed method are similar to those of the real-time mathematical programming model, the computational time is reduced by over 95\% due to the adoption of model decomposition and the distributed computing approach. This makes the computational time of the proposed method comparable to that of the single deterministic model. Additionally, its profit increased by 23.61\% compared to the deterministic model. Although the profit is 6.81\% lower than that of the real-time optimization-based mathematical programming model, this is primarily due to the challenge of fully considering global constraints and optimization in the distributed optimization method. However, the performance is expected to improve gradually with an increase in the number of training iterations. Nevertheless, the proposed method significantly enhances the refinery’s ability to respond to external product price fluctuations.

\begin{table}[ht]
\centering
\small  % 设置表格字体为小
\caption{Performance Comparison of The Three methods}
\begin{tabular}{|c|c|c|c|}

\hline
\textbf{Performance } & \textbf{Deterministic MP model} & \textbf{Our method} & \textbf{Real-time MP model} \\
\hline
\textbf{Solution} & 108660.16 & 134319.38 & 144143.30 \\
\hline
\textbf{Total Time(s)} & 4.59 & 22.38 & 1022.97 \\
\hline
\textbf{Number of Equations} & 551 & 54600 & 55100 \\
\hline
\textbf{Number of Variables} & 797 & 79200 & 79700 \\
\hline
\textbf{Number of Binary variables} & 11 & 1100 & 1100 \\
\hline
\end{tabular}

\label{The Performance Comparision of The Three Methods in Case 1}
\end{table}

\subsection{Simple Multi-period refinery planning case}
The method we proposed is not only applicable to the single-period model but can also be effectively adapted to the multi-period refinery planning model. To demonstrate this, we adopted a simple multi-period refinery planning linear programming (LP) model from H. P. Williams \cite{williams2013model} to evaluate the performance of our method in a multi-period refinery planning context.

\begin{figure}
    \centering
    \includegraphics[width=0.75\linewidth]{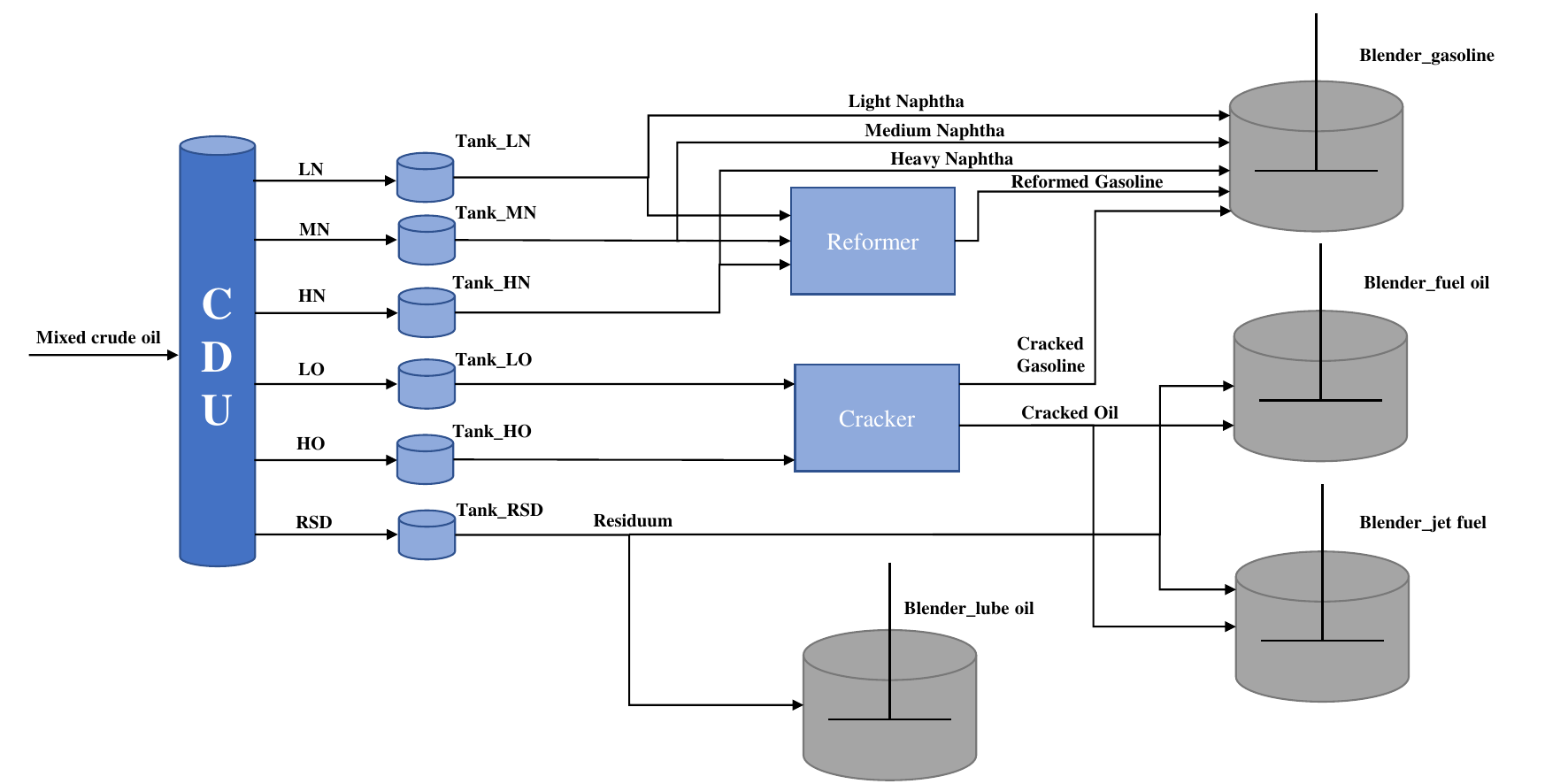}
    \caption{The refinery flowchart in Case 2}
    \label{fig:case2}
\end{figure}

The refinery planning network and its decomposition method are shown in Figure \ref{fig:case2}. The refinery can purchase two types of crude oil with different qualities. After blending the crude oils evenly, they are sent to CDU for distillation. The outlet streams from the CDU are stored in tanks until they are required for further processing. Some of the intermediate product streams are then processed through reforming and cracking units to enhance quality. Finally, all intermediate products are blended in blenders to obtain four products: gasoline, fuel oil, jet fuel, and lubricants, which are subsequently sold in the market for profit.

\begin{figure}
    \centering
    \includegraphics[width=0.75\linewidth]{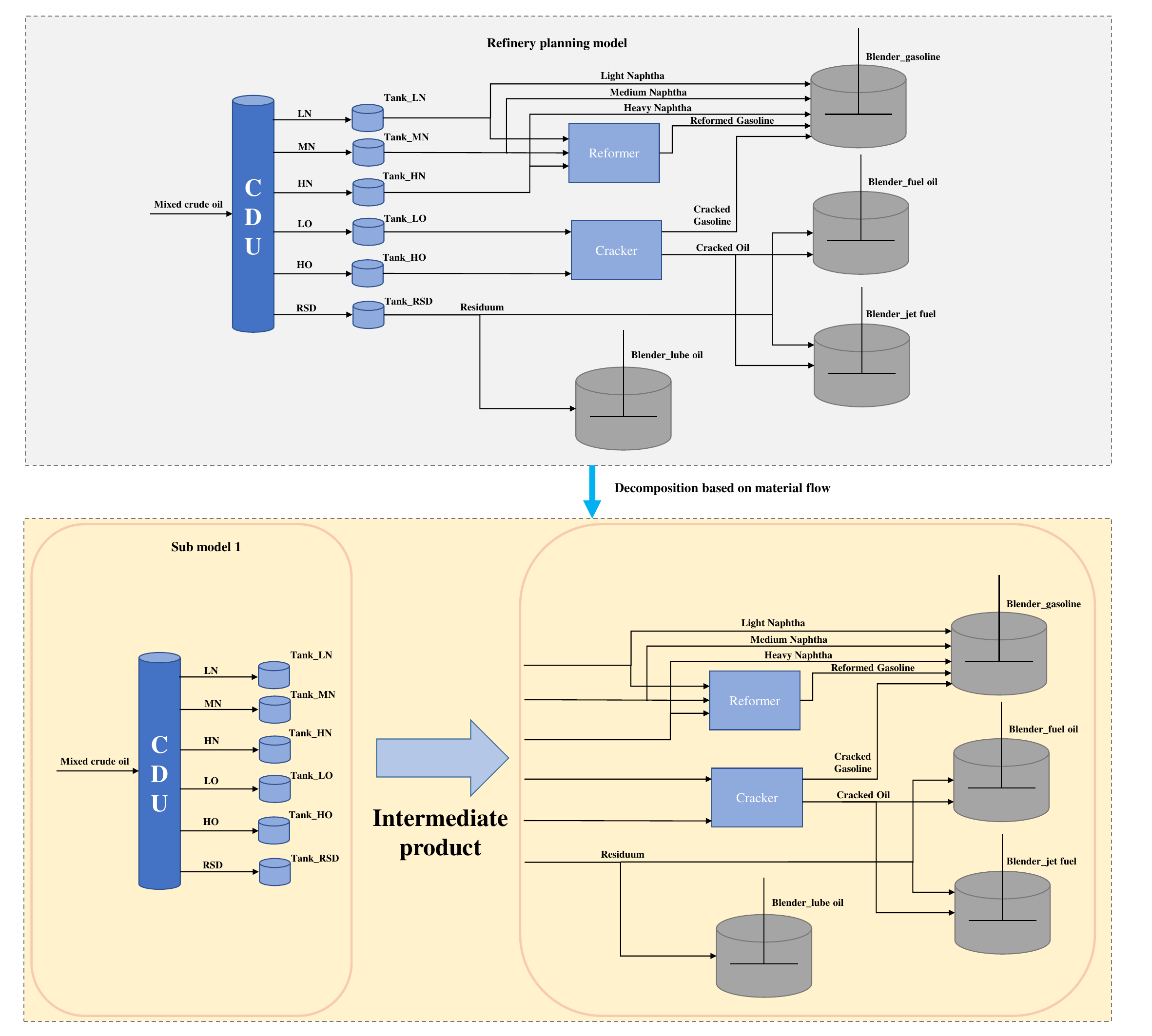}
    \caption{The decomposition method for Case 2}
    \label{fig:decomposition case2}
\end{figure}

By applying the decomposition method, the refinery model is decomposed into two sub models, as shown in Figure \ref{fig:decomposition case2}. The objective functions for sub model 1 and 2 are defined as follows:

\begin{equation}
    Obj_{1,t}=F(\omega_{1,t},\lambda_{1,t})=\sum_{s:(u,s)\in UO}P_{s,t}\cdot VS_{s,u,t}\ \ \ \ \forall t\in T,u\in CDU
\end{equation}
where there is no term related to the quality of oil cuts because the quality parameters in this model are fixed.
\begin{equation}
    Obj_{2,t}=F(\omega_{2,t},\lambda_{2,t})=\sum_{p\in P}P_{p,t}\cdot VP_{p,t}-\sum_{s:(u,s)\in IO}P'_{s,t}\cdot VI_{s,i,t}\ \ \ \forall t\in T,i\in I
\end{equation}
where $P'_s$ can be understood as the price of the intermediate product, which is influenced by supply and demand. Therefore, its specific value is related to the current inventory level of the product.
\begin{equation}
    P'_{s,t}=P_{s,t}-\alpha\cdot I_{s,t} \ \ \ \ \forall t\in T,i \in I
\end{equation}
where $\alpha$ is a hyperparameter and is set to $1.5*10^{-5}$ in this case.

To ensure planning responsiveness to price fluctuations, it is essential to integrate market price changes into the planning model. However, the number of scenarios increases exponentially with the number of time steps. A practical way is to formulate this problem as a dynamic optimization problem and solve it in a receding horizon. We assume that the refinery can observe future price fluctuations in a probabilistic manner and thus formulate multi-period scenery-based MILP model as follows. 
\begin{equation}
    Obj_t=\sum_{p\in P}P_{p,t}\cdot VP_{p,t}+\sum_{sce\in SCE}\sum_{p\in P}P_{sce,p,t+1}\cdot VP_{sce,p,t+1}\ \ \ \ \forall t\in T
\end{equation}
\begin{equation}
    P_{sce,p,t+1}\sim \mathcal{N}(\mu_{sce,p,t+1},\sigma^2_{sce,p,t+1})\ \ \ \ \ \forall p\in P,t\in T,sce \in SCE
\end{equation}
where $P_{sce,p,t+1}$ is the product $p$ price sampling in the scenerio $sce$ at time step $t+1$ and the number of total scenery at time $t+1$ is set to 10.
\begin{equation}
    Profit = \sum_{t\in T}\sum_{p\in P}P_{p,t}\cdot VP_{p,t}
\end{equation}

The economic data from March 2017 to April 2022 (approximately 230 data points) from the U.S. Energy Information Administration was used.

In this case, the state space, action space, and reward are defined as follows: i.e.$\mathscr{S}\triangleq[P_{p,t},\hat{P}_{p,t+1},I_{i,t}]$,$\mathscr{A}\triangleq[P_{s,t}]$,$R=Profit$.

\begin{figure}
    \centering
    \includegraphics[width=0.8\linewidth]{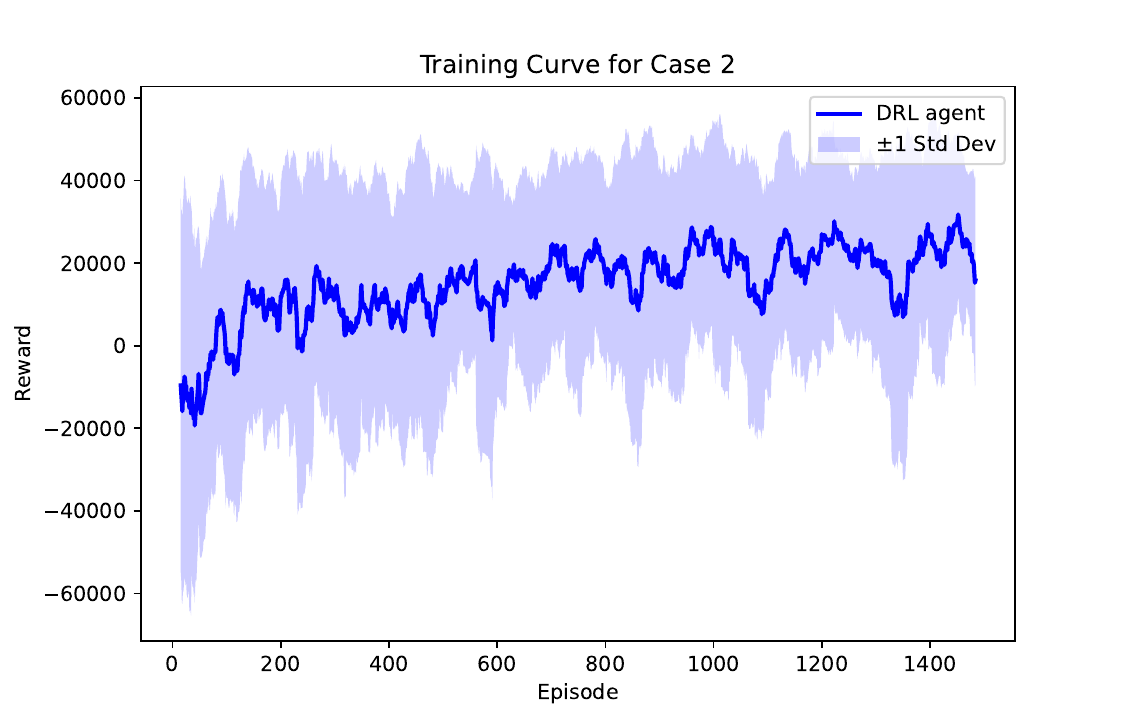}
    \caption{The training curve of pricing strategy in Case 2}
    \label{fig:training curve case2}
\end{figure}

The relevant parameters of the pricing strategy are set the same as in Case 1 except the number of training episodes and the batch size, which are set to 1500 and 60, respectively. Training the pricing strategy for this case takes about 13 minutes. Figure \ref{fig:training curve case2} visualizes the training curve of this case. After 650 training iterations, although some local fluctuations remain, the agent’s reward stabilizes at a consistently positive value. The reason for the fluctuations is that the actual future prices are sampled from a distribution, leading to some deviation from the expected values.

\begin{figure}
    \centering
    \includegraphics[width=0.5\linewidth]{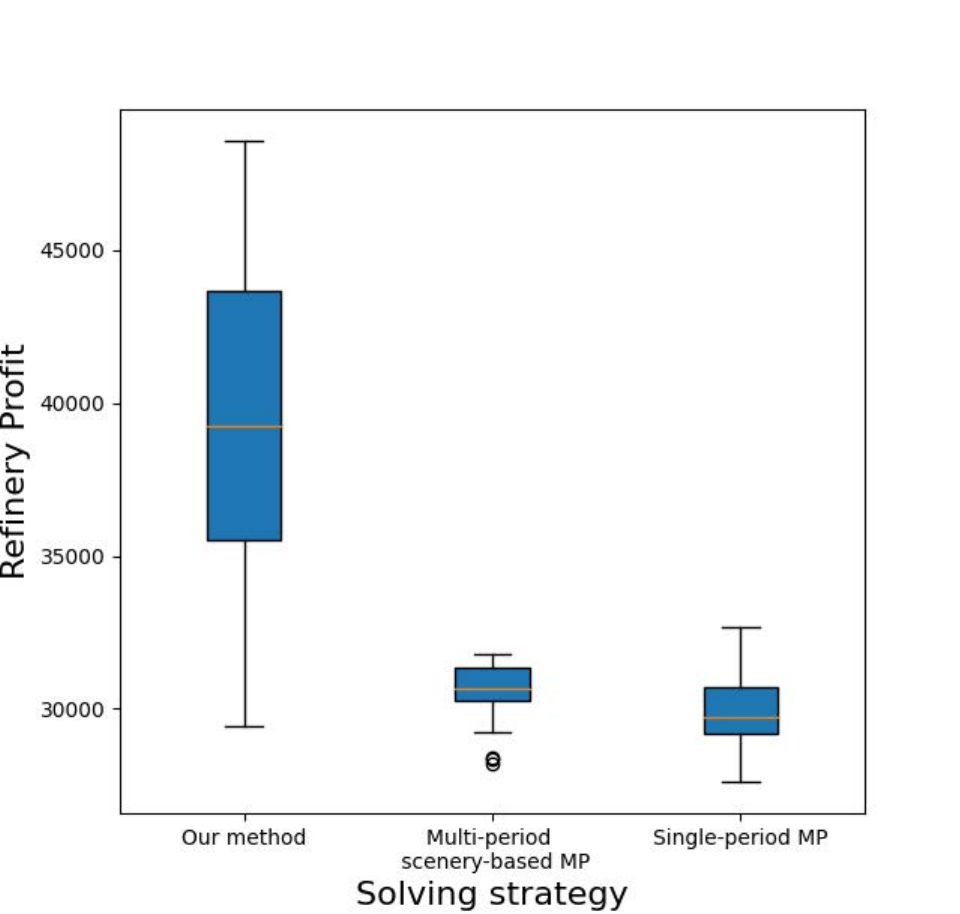}
    \caption{The box plot of refinery planning profit with three methods across 20 runs}
    \label{fig:profit case2}
\end{figure}

We compare the performance of the method we proposed with a multi-period scenery-based LP planning model and a single-period LP planning model. Figure \ref{fig:profit case2} presents a box plot of refinery profits earned by the multi-period planning model with different strategies over 60 days for 20 runs. The average profits for the three planning strategies are 38809.95, 30506.89 and 29936.51 respectively. From the figure, it is evident that our method has a better performance than the other strategies with an improvement of 27.2\% and 29.6\%, respectively. This superior performance can be attributed to the following reason: while both our method and the scenario-based multi-period mathematical programming model can obtain the same information, the pricing strategy based on DRL in our method is capable of gaining insights into price fluctuations during the training process and ensures the acquisition of high-quality solutions. The summarized performance comparison is presented in Table \ref{Performance Comparision of three method case2}.

\begin{figure}
    \centering
    \includegraphics[width=0.75\linewidth]{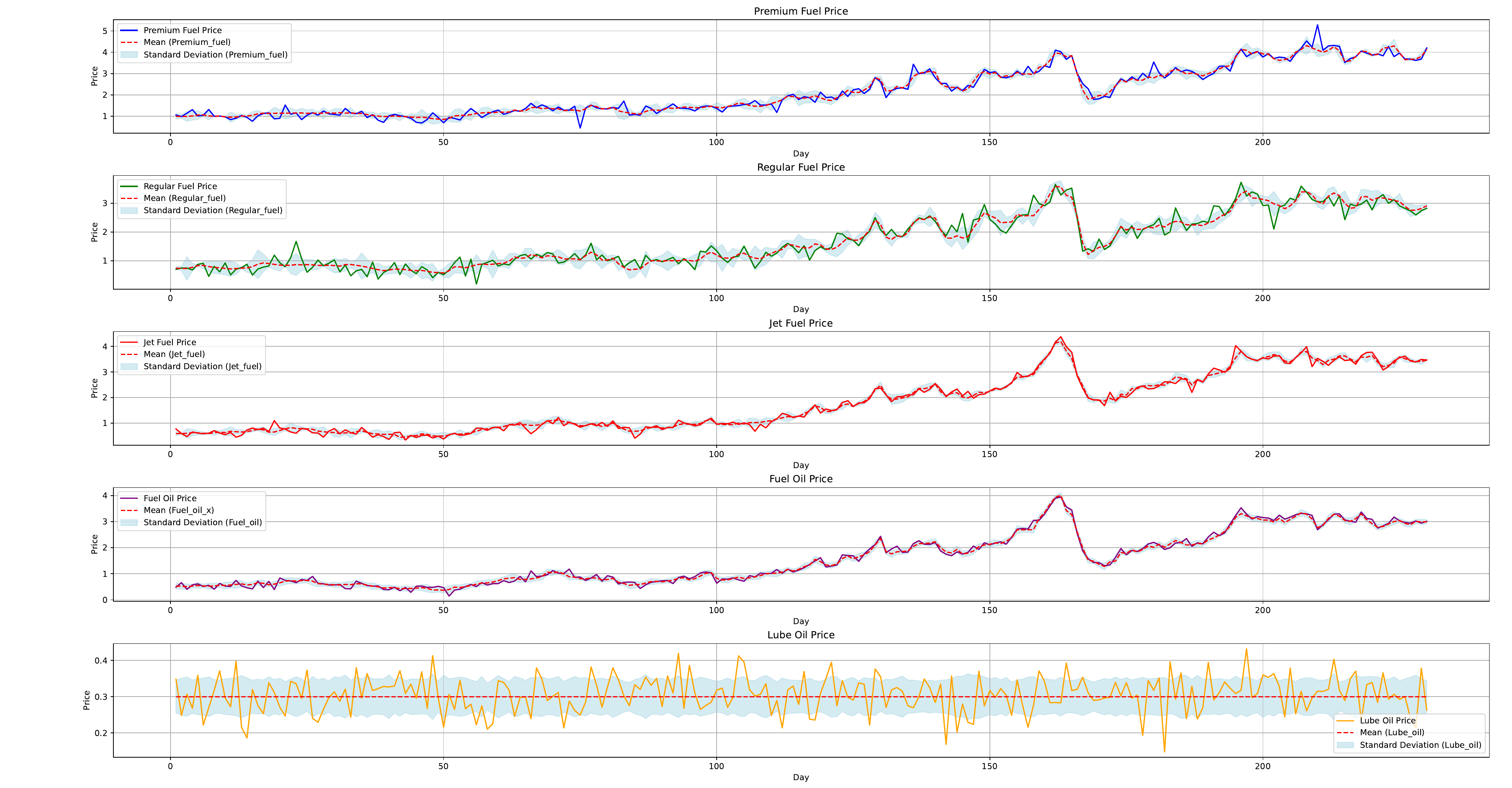}
    \caption{The price trends of petroleum products in Case 2}
    \label{fig:price case2}
\end{figure}

\begin{table}[ht]
\centering
\small  % 设置表格字体为小
\caption{Average Performance Comparison of Three method}
\begin{tabular}{|c|c|c|c|}
\hline
\textbf{Performance} & \textbf{Our method} & \textbf{\shortstack{Multi-period\\scenery-based MP}} & \textbf{Single-period model} \\
\hline
\textbf{Solution(\$)} & 38809.95 & 30506.89 & 29936.51 \\
\hline
\textbf{Total Time(s)} & 6.375 & 10.662 & 1.459 \\
\hline
\textbf{Number of Constraints} & 2820 & 25020 & 1740 \\
\hline
\textbf{Number of Variables} & 3600 & 32040 & 2160 \\

\hline
\end{tabular}

\label{The Average Performance of The Three Methods in Case 2}
\end{table}

\subsection{Multi-period refinery planning case with MINLP model}

\begin{figure}
    \centering
    \includegraphics[width=0.75\linewidth]{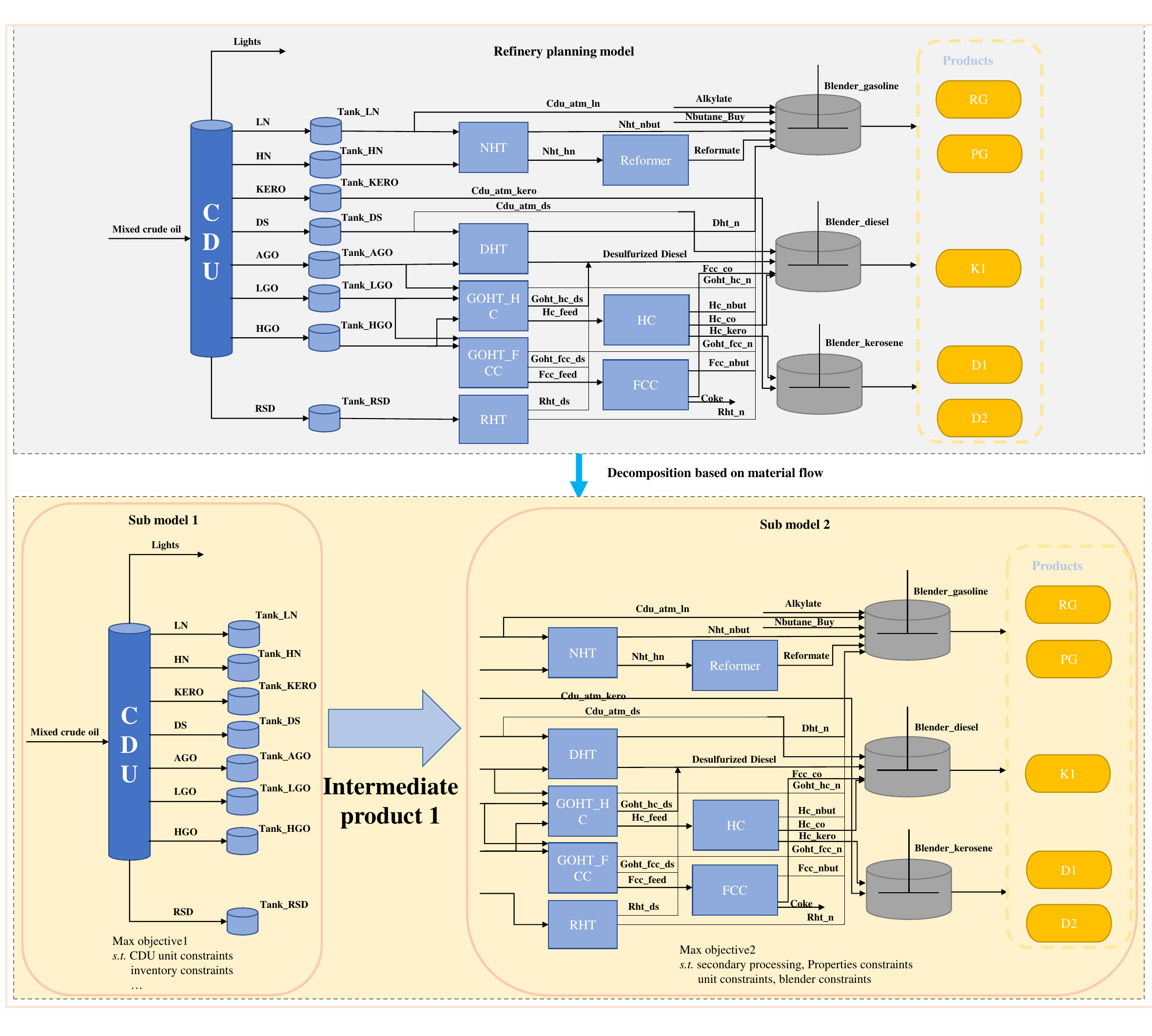}
    \caption{The Decomposition method for Case 3}
    \label{fig:case3}
\end{figure}

We also studied the performance of the proposed method in the multi-period large-scale complexly coupled refinery planning problem, as illustrated in \ref{fig:case3}. The only difference from the refinery network in Case 1 is the addition of storage tanks, capable of storing intermediate products. After decomposing, the objective function designed for sub model 1 is the same as $Eq(58)$, while the $Obj_2$ should take inventory cost into account:
\begin{equation}
\begin{split}
    Obj_{2,t}=F(\omega_{2,t},\lambda_{2,t})=\sum_{t\in T}(\sum_{p\in P}P_{p,t}\cdot VP_{p,t}\ -\ \sum_{c\in C}P_{c}\cdot VC_{c,t} -\\
    \ \sum_{u\in U}P_{u}\cdot VU_{u,t}\ -\sum_{m\in M}P_m\cdot VS_{m,t}\ -\sum_{s:(i,s)\in IO}P'_{s,t}\cdot VI_{i,t})
\end{split}
\end{equation}
where $P'_{i,t}$ incorporates the inventory cost, and dynamically adjusts the production rates in the subsequent stages based on inventory levels.
\begin{equation}
    P'_{s,t}=P_{s,t}-\alpha\cdot I_{i,t}\ \ \ \forall i\in I,s:(i,s)\in IO,t\in T
\end{equation}

For pricing strategy based on DRL of this case, the state space, the action space and reward are defined as $\mathscr{S}\triangleq[P_{p,t},\hat{P}_{p,t+1},I_{i,t-1},QI_{i,q,t-1}]$, $\mathscr{A}\triangleq[P_{s,t},P_{s,q}]$ and $R=Profit$, respectively. This means that the pricing strategy will make decisions based on the current market price, expected future market price, the current inventory levels, and the quality properties of inventory products. It then dynamically adjust the sub models' pricing parameters to guide the refinery planning towards profit maximization.
\begin{equation}
    Profit=\sum_{t\in T}(\sum_{p\in P}P_{p,t}\cdot VP_{p,t}\ -\ \sum_{c\in C}P_{c}\cdot VC_{c,t} -
    \ \sum_{u\in U}P_{u}\cdot VU_{u,t}\ -\sum_{m\in M}P_m\cdot VS_{m,t} - \sum_{i\in tanks}C_i\cdot I_{i,t})
\end{equation}

\begin{figure}
    \centering
    \includegraphics[width=0.8\linewidth]{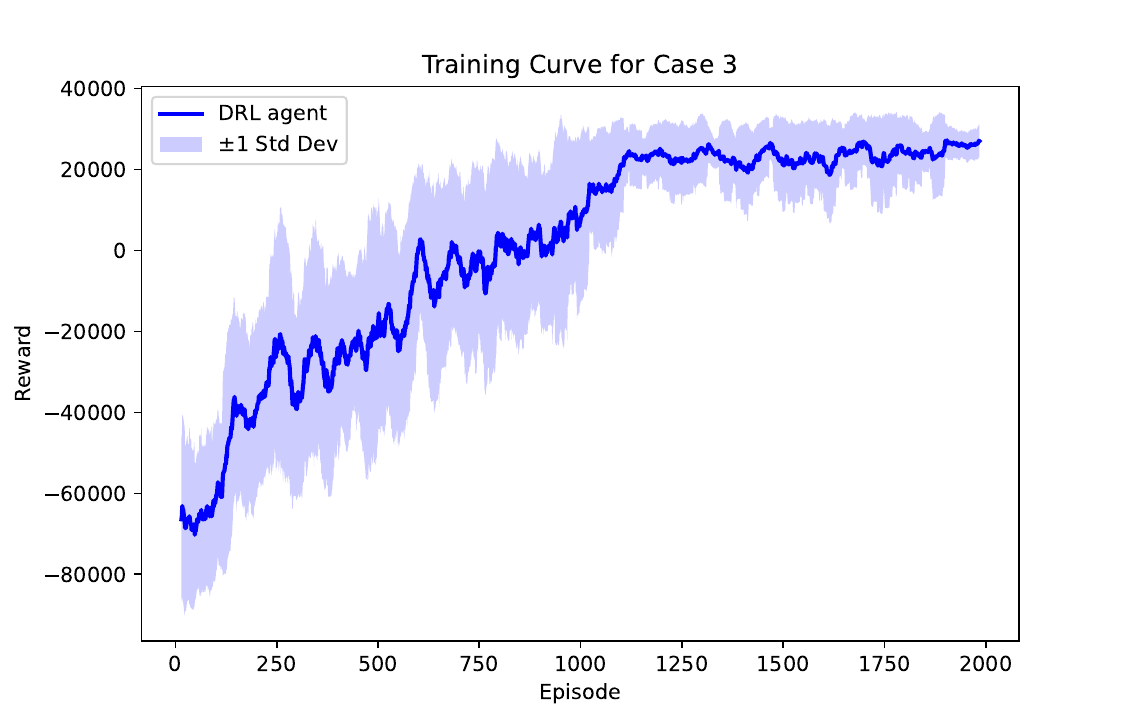}
    \caption{Reward for pricing strategy in Case 3 during 2000 episodes}
    \label{fig:training_cureve_case3}
\end{figure}

 The hyperparameters are the same as those in Case 1. After several training iterations and parameter tuning, the reward for pricing strategy steadily increased, as shown in Figure \ref{fig:training_cureve_case3}. Due to the inherent complexity of the non-convex MINLP planning model, even after decomposition, the pricing strategy still requires a prolonged training period (about 4.82h) to achieve good performance.

\begin{figure}
    \centering
    \includegraphics[width=0.8\linewidth]{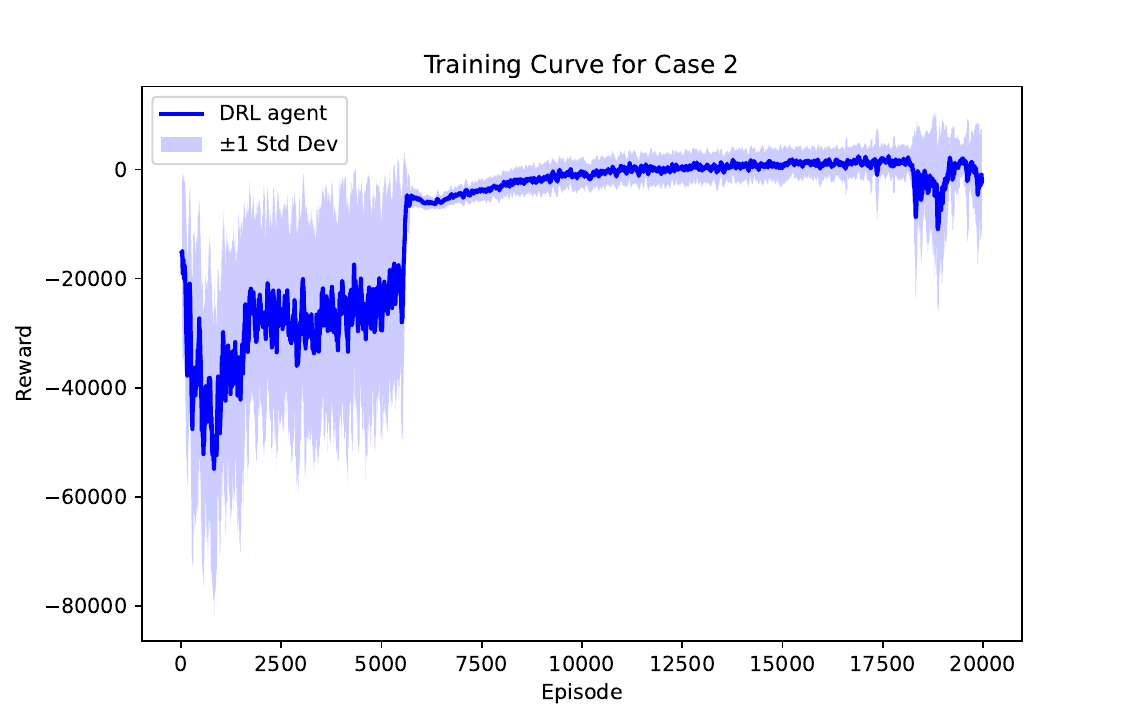}
    \caption{Reward of direct RL optimization during 20000 episodes}
    \label{fig:direct_RL}
\end{figure}

We employ the multi-period MINLP model solved by mathematical programming method to be our baseline, which consists of the objective function shown in $Eq(70)$, the processing unit constraints in $Eq(7)-(34)$, inventory constraints in $Eq(35)-(41)$ and blender constraints in $Eq(42)-(48)$. 

Another experimental benchmark involves directly applying DRL to this refinery planning problem. For this benchmark experiment, the state space and reward function are the same as those of pricing strategy in Case 3, while the action space consists of the volumes of each type of crude oil to import and the storage level of each tank, i.e. $\mathscr{A}\triangleq[VC_{c,t},I_{i,t}]$. Due to the existence of non-linear blending formulas, the action space cannot specify the storage level of each individual tank directly. Instead, it optimizes by specifying the storage levels of certain key variables. In this experimental benchmark, the intermediate products whose storage levels can be specified are $LN$, $KERO$ and $DS$. The hyperparameter settings are consistent with those of pricing strategy in  Case 3, with the only difference being the training iterations set to 20,000.

In Case 3, the refinery should make production planning strategy for four periods, i.e. $T=4$. For the optimization of mathematical programming, we set the termination condition as when the time reaches a specified threshold or the gap between the upper and lower bounds narrows to a certain degree, i.e.$T=60s\ or \ MIPgap = 0.001$, the Gurobi solver outputs the current solution as the optimal solution.

\begin{figure}
    \centering
    \includegraphics[width=0.5\linewidth]{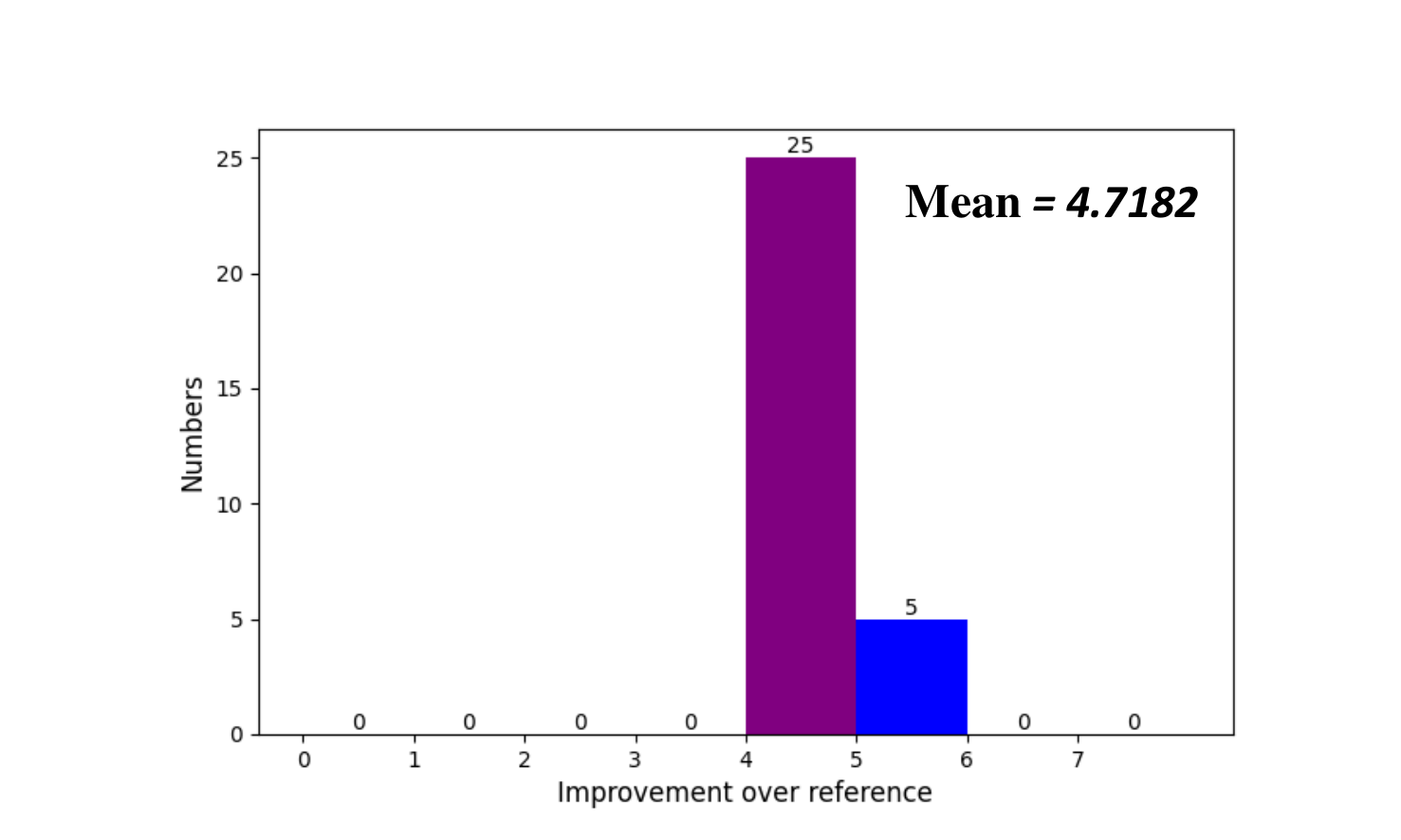}
    \caption{Histogram of the sampled improvement of the proposed method over the multi-period MINLP model across 30 runs}
    \label{fig:improvement over}
\end{figure}

\begin{figure}
    \centering
    \includegraphics[width=0.75\linewidth]{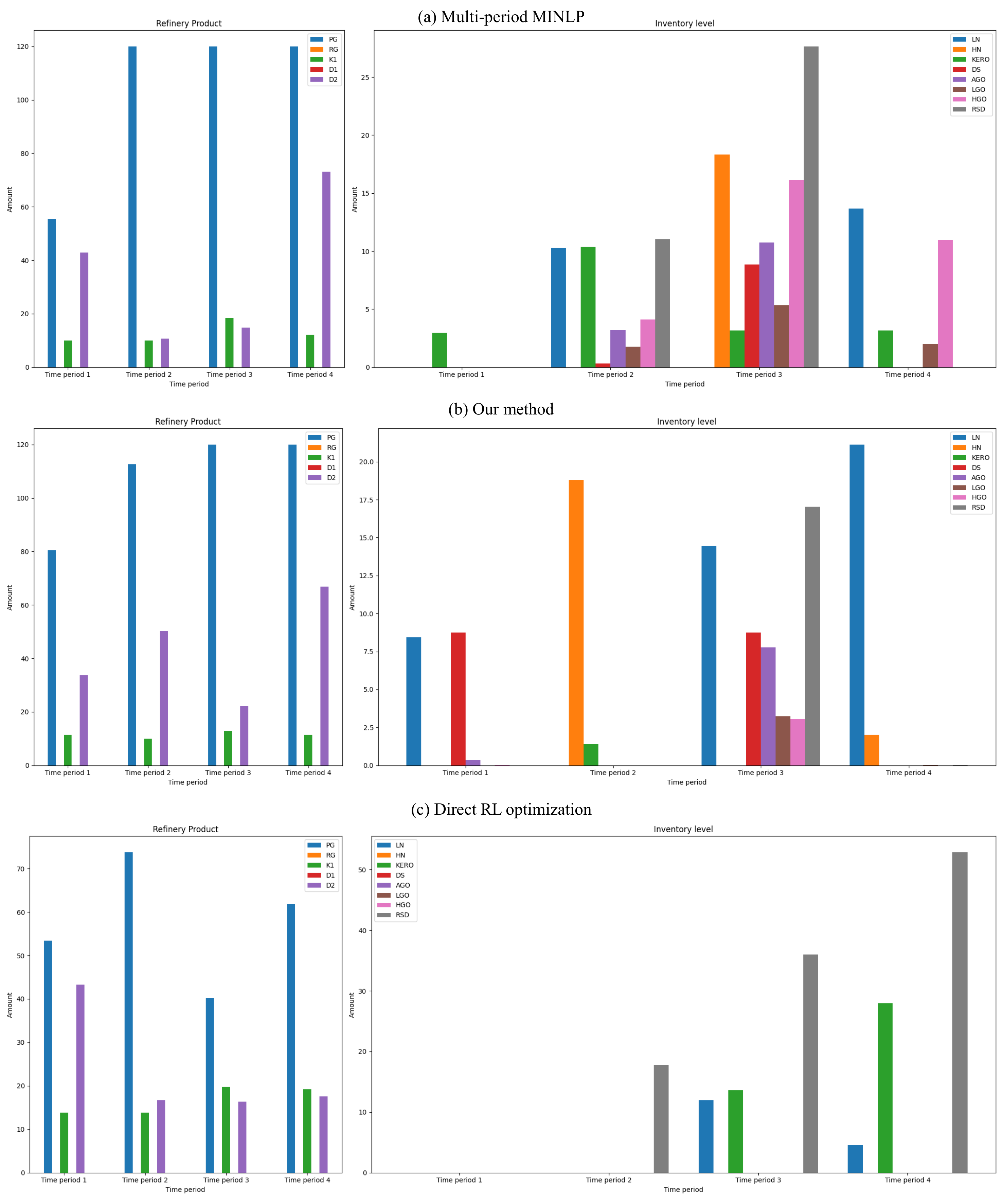}
    \caption{Comparison of the average inventory level and product produced during each time period based on the three methods}
    \label{fig:product and inventory}
\end{figure}

In Figure \ref{fig:product and inventory}, the left side shows the average product produced during each period when applying the three methods, while the right side shows the average inventory level of intermediate product stored in the tanks. 

Figure \ref{fig:improvement over} displays the distribution of the improvement achieved by the proposed policy (training after 1000 episodes and 0.2 exploration rate) compared to the reference multi-period MINLP method across 30 runs. From the results of this case, it is evident that the proposed method outperforms all others, consistently surpassing the reference multi-period MINLP mathematical programming method in all 30 trials, with an average improvement of 4.7182\%. The complete comparative results of the three methods are shown in Table \ref{Performance Comparision of three method case3}. 

It can be observed that by applying our method, the refinery earns the highest profit. This is partially because the reinforcement learning-based training approach enables the pricing strategy to learn predictive information about future prices, allowing for decision-making that outperforms methods based solely on multi-period mathematical programming. The method we proposed ensures profit while significantly improving solving speed, with a gap of only 1.9079\% compared to 4.9996\% for the multi-period mathematical programming method. Additionally, the computation time decreased from 241.864 $s$ to 146.973 $s$, demonstrating a significant increase in computational efficiency.

In addition, from the results in Figure \ref{fig:direct_RL} and Table \ref{Performance Comparision of three method case3}, we observe that even after 20,000 training iterations (about 3.64h), the performance of the RL agent in directly formulating the multi-period large-scale refinery plan concerning crude oil imports and inventory remains unsatisfactory. The trained solutions tend to get trapped in local optima, with an average reward of only 4593.78, and exhibit divergence as the number of training iterations increases. The bar chart in the Figure \ref{fig:product and inventory} shows that directly applying the RL method to the production planning of the large-scale non-convex refinery leads to a significant reduction in refinery profit. This approach results in lower product output per cycle compared to the first two methods, while also complicating the storage of valuable intermediate products in alignment with price trends for future production and sales. These results suggest that adopting the strategy based on directly RL optimization for refinery planning could result in infeasible solutions, significantly reducing the feasibility of directly applying RL to large-scale non-convex refinery planning optimization.

\begin{table}[ht]
\centering
\small  % 设置表格字体为小
\caption{The Performance Comparison of The Three method in Case 3}
\begin{tabular}{|c|c|c|c|}

\hline
\textbf{Performance } & \textbf{Multi-period MP} & \textbf{Our method} & \textbf{Direct RL optimization} \\
\hline
\textbf{Mean Reward} & 24016.0938 & 25149.2250 & 4438.2210 \\
\hline
\textbf{Total Time(s)} & 241.864 & 146.973 & 0.355 \\
\hline
\textbf{Gap(\%)} & 4.9996 & 1.9079 & 0.0001 \\
\hline
\end{tabular}

\label{Performance Comparision of three method case3}
\end{table}

\section{Conclusion}\label{Conclusion}

This paper proposed an innovative optimization framework for large-scale complexly coupled refinery planning problems, leveraging model decomposition and pricing strategy based on DRL. Initially, the refinery is decomposed into multiple sub models based on the logistical relationships and physical material flow within the refinery network. Each sub model contains a subset of the refinery's planning decision variables and corresponding constraints. Subsequently, an appropriate objective function is defined for each sub model, and a reinforcement learning-based pricing strategy is proposed to dynamically assign function parameters. Finally, the overall refinery planning strategy is derived by sequentially solving the sub models using mathematical programming methods, following the logistics order.

The proposed method is highly interpretable, significantly improving computational efficiency, and validly addressing the impact of price uncertainty. Three real-world industrial cases, including both single-period and multi-period planning with simple MILP and complex MINLP refinery planning models, demonstrate the feasibility and effectiveness of the method. The results demonstrate that the method proposed in this paper can greatly enhance the solution speed and convergence efficiency of the plant-wide refinery planning problem. Additionally, by utilizing the pricing strategy based on DRL, the dynamic characteristics of price fluctuations are captured, enabling the development of a more robust refinery management planning strategy.
 
Future work will focus on investigating the method's performance under more complex uncertain conditions and developing a more flexible and general framework for building sub models, which will further expand the applicability and scope of the method.

\section*{Acknowledgement}

\section*{Nomenclature}

\subsection*{Sets and Indices}

\begin{longtable}{@{} >{\ttfamily}l @{\ :\ } p{0.8\textwidth} @{}}  
  $C=\{c\}$       & Set of crude oils \\
  $U=\{u\}$       & Set of units \\
  $S=\{s\}$       & Set of intermediate product streams \\
  $P=\{p\}$       & Set of final products \\
  $I=\{i\}$       & Set of inventory tanks \\
  $B=\{b\}$       & Set of blenders \\
  $Q=\{q\}$       & Set of quality properties \\
  $T=\{t\}$       & Set of time periods \\

\end{longtable}

\subsection*{Subsets}

\begin{longtable}{@{} >{\ttfamily}l @{\ :\ } p{0.8\textwidth} @{}}  
  $BI = \lbrace (b, s) \rbrace$ & Inlet streams of blender $b$ \\
  $BO = \lbrace (b, s) \rbrace$ & Outlet streams of blender $b$ \\
  $II = \lbrace (i, s) \rbrace$ & Inlet streams of storage tank $i$ \\
  $IO = \lbrace (i, s) \rbrace$ & Outlet streams of storage tank $i$ \\
  $UI = \lbrace (u, s) \rbrace$ & Inlet streams of unit $u$ \\
  $UO = \lbrace (u, s) \rbrace$ & Outlet streams of unit $u$ \\
  $CDU = \lbrace u \rbrace$ & Crude distillation units \\
  $HTU = \lbrace u \rbrace$ & Hydrotreating units \\
  $MMU = \lbrace (ru,u) \rbrace$ & unit $u$ that represent one operating mode of a real unit $ru$ with multiple operating modes \\
  $MU = \lbrace u \rbrace$ & Real units with multiple operating modes \\
  $SU = \lbrace u \rbrace$ & Real units with single operating modes \\
  $SQV = \lbrace q \rbrace$ & Quality properties that blend linearly on a volumetric basis \\
  $SQW = \lbrace q \rbrace$ & Quality properties that blend linearly on a weight basis \\
  
\end{longtable}

\subsection*{Parameters}
\begin{longtable}{@{} >{\ttfamily}l @{\ :\ } p{0.8\textwidth} @{}}  
  $P_{p,t}$       & Market price for product $p$ during period $t$ \\
  $UMM_{ru,t}$       &  maximum number of modes in which real unit $ru$
 can operate during a time period $t$\\
  $V^{min}_{u,t}$       & Minimum volume to be processed by unit $u$ during period $t$ \\
  $V^{max}_{u,t}$       & Maximum volume to be processed by unit $u$ during period $t$ \\
  $V^{min}_{ru,t}$       & Minimum volume to be processed by a real unit $ru$ during period $t$ \\
  $V^{max}_{ru,t}$       & Maximum volume to be processed by a real unit $ru$ during period $t$ \\
  $Yield_{u,c,s}$       & Yield of oil cut $s$ from crude oil $c$ by CDU $u$\\
  $QC^{fix}_{q,c,s}$       & Fixed quality value of property $q$ in distillation cut $s$ when feeding CDU with crude oil $c$\\
  $QC^{fix}_{'sg',c,s}$       & Fixed quality value of property specific gravity in distillation cut $s$ when feeding CDU with crude oil $c$\\
  $Yield^{fix}_{u,s}$       & Fixed yield of stream $s$ from unit $u$\\
  $QS^{fix}_{s,q,t}$       &  Fixed quality value of property $q$  in stream $s$ during period $t$\\
  $RSUL^{max}_{u,t}$       &  The upper limit of sulfur removal capacity of unit $u$ during period $t$\\
  $HTU_u^{max}$       &  Maximum sulfur remove rate of unit $u$ during period $t$\\
  $HTU_u^{min}$       &  Minimum sulfur remove rate of unit $u$ during period $t$\\
  $SulR_{u,t}^{fix}$       & Fixed sulfur content remove rate of unit $u$ during period $t$\\
  $I_i^{ini}$       & Initial inventory level of tank $i$\\
  $IL_i^{min}$       & Minimum inventory level of inventory tank $i$\\
  $IL_i^{max}$       & Maximum inventory level of inventory tank $i$\\
  $QI_{i,q}^{ini}$       & Initial quality property $q$ of intermediate product stored in tank $i$\\
  $QI_{i,'sg'}^{ini}$       & Initial specific gravity of intermediate product stored in tank $i$\\
  $VB_{b,t}^{min}$       & Minimum volume to be blended by the blender $b$ during period $t$ \\
  $VB_{b,t}^{max}$       & Maximum volume to be blended by the blender $b$ during period $t$\\
  $QP^L_{p,q,t}$       & Lower bound of quality property $q$ of product $p$ during period $t$\\
  $QP^U_{p,q,t}$       & Upper bound of quality property $q$ of product $p$ during period $t$\\
  % 添加更多条目... 
\end{longtable}

\subsection*{Binary Variables}
\begin{longtable}{@{} >{\ttfamily}l @{\ :\ } p{0.8\textwidth} @{}}  
  $bu_{u,t}$       & If equal to 1, then unit $u$ which is a operating mode of a real unit $ru$ operates during time period $t$ \\
  $bb_{p,t}$       & If equal to 1, then the blender will produce product $p$ during period $t$ \\
  
\end{longtable}

\subsection*{Continuous Variables}
\begin{longtable}{@{} >{\ttfamily}l @{\ :\ } p{0.8\textwidth} @{}}  
  $VS_{s,u,t}$       & Volume of stream $s$ through unit $u$ during period $t$ \\
  $QS_{s,q,t}$       & Quality property $q$ of stream $s$ during period $t$ \\
  $I_{i,t}$       & Inventory level of intermediate product $i$ during period $t$ \\
  $VU_{u,t}$       & Volume of total stream flow through unit $u$ during period $t$ \\
  $VU_{s,ru,t}$       & Volume of stream $s$ through a real unit $ru$ with multiple operating modes during period $t$ \\
  $VU_{ru,t}$       & Volume of total stream flow through a real unit $ru$ with multiple operating modes during period $t$ \\
  $VC_{c,t}$       & Volume of crude oil $c$ processed in CDU during period $t$ \\
  $QS_{u,'sg',t}$       & Quality value of property specific gravity in the total feed stream through unit $u$ during period $t$\\
  $WS_{s,u,t}$       & Weight of stream $s$ which flow through unit $u$ during period $t$\\
  $QS_{s,'sg',t}$       & Quality value of property specific gravity of stream $s$ during period $t$\\
  $SulS_{s,u,t}$       & Total sulfur content of stream $s$ which flow through unit $u$ during period $t$\\
  $QS_{s,'sul',t}$       & Quality value of property sulfur content of stream $s$ during period $t$\\
  $VI_{s,i,t}$       & Volume of stream $s$ through inventory tank $i$ during period $t$  \\
  $I_{i,t}$       & Inventory level in tank $i$ at the end of period $t$  \\
  $QI_{i,q,t}$       & Quality value of property $q$ of intermediate product stored in tank $i$ at the end of period $t$  \\
  $QI_{s,'sg',t}$       & Specific gravity of intermediate product stored in tank $i$ at the end of period $t$\\
  $VB_{b,t}$       & Volume of total stream flow into blender $b$ during period $t$  \\
  $VS_{s,b,t}$       & Volume of stream $s$ through blender $b$ during period $t$ \\
  $VP_{p,t}$       & Volume of product $p$ produced during period $t$ \\
  $QP_{p,q,t}$       & Quality value pf property $q$ of product $p$ during time period $t$\\
  $QP_{p,'sg',t}$       & Specific gravity of product $p$ during time period $t$\\
  
\end{longtable}

\section*{References}
\bibliographystyle{vancouver}
\bibliography{refs}

\appendix

\section{An Explanation for Objective Function of Sub Models }\label{Provement}

As stated in Section\ref{Basis}, for a planning problem, there must exist an optimal solution, which we will denote as $\omega^*$ representing all the decision variables. It is evident that the optimal solution can be divided into different sections according to different unit, for example, we can name the decision variables of CDU unit as $\omega_1$ which including the inflow rate, blend recipe of crude oil, flow rate and quality properties of oil cuts, $\omega_2$ represents the variables of inventory tanks, which including the inventory level of intermediate product, the quality properties of inventory, and $\omega_3$ may be the operating modes, inflow streams from tanks to the secondary units, flow rate and quality properties of outlet stream of secondary units, $\omega_4$ is the corresponding variables for blenders i.e. 

And we follow the approach that, after decomposing the refinery into multiple sub models based on its material flows, the optimal solution can be got by solving the sub models by sequence as long as we can design proper objective function for them.

That means we should make sure the solution of sub model is optimal when $\omega_i =\omega_i^*$, where $\omega_i^*$ is the decision variables for sub model as a part of optimal solution of refinery planning model. However, the optimal solution may change as the product price changed. Therefore, we try to design the general form of objective function of sub models under price uncertainty rather than a specific function for a deterministic problem.

Many studies on refinery planning have adopted the fixed-yield model for CDU so as ours. And the general form of CDU's objective function can be 
\begin{equation}
    Obj_i=\sum_{s:(u,s)\in UO}(P_s\cdot VS_s +\sum_{q\in Q} P_{s,q}\cdot QS_{s,q})+P^{'}_{u}VU^2_{u}\ \ \ \ \forall u\in CDU
    \label{Obj1}
\end{equation}
We can use the vector $F_C=[C_1,C_2,...,C_m]^T$ to represent the imported crude oil and $F_S=[S_1,S_2,...S_n]^T$ to represent the oil cut with quality properties $Q_s=[Q_{sv},Q_{sg}]^T=[Q_{1,1},Q_{1,2},...Q_{n,v},Q_{n,g}]^T$ where $Q_{sv}$ and $Q{sg}$ are quality based on volume or gravity, respectively. As we employ the fixed-yield model, it is evident that
\begin{equation}
    F^T_S = A_1\cdot F_C
\end{equation}

Let us assume that the quality of each oil cuts is represented by $Q_{iv}$ and $Q_{isg}$, corresponding to quality based on flow rate and specific gravity, respectively. In the formula of the fixed output model, these qualities can be expressed as the product of the corresponding terms for crude oil, that is:

\begin{equation}
    Q_{iv}\cdot S_i\ =\ A_{2,i}\cdot F_C
\end{equation}

\begin{equation}
    Q_{v}\cdot F^T_S = A_2 \cdot F_C 
\end{equation}
where $Q_v=[Q_{1v},Q_{2v},...Q_{nv}]$,$A_2=[A_{2,1},A_{2,2},...A_{2,n}]$.
Similarly, the quality properties of oil cuts based on gravity can be expressed as:
\begin{equation}
    Q_{sg}\cdot Q_{s,'sg'}\cdot F^T_s = A_3 \cdot F_c
\end{equation}
where $A_1$, $A_2$, $A_3$ are formed by the combination of parameters in the fixed-yield model. 

From the above equations, it can be observed that the CDU outlet flow rate and the product of the outlet flow rate and quality can both be expressed as polynomials of the crude oil feed. If
$m$ linearly independent terms can be selected from the coefficient matrix $A = \begin{pmatrix}
A_1 \\
A_2 \\
A_3
\end{pmatrix}$, all other terms can be expressed as linear combinations of these terms.

So that we can modify the objective function form according to the rank of parameter matrix of the CDU fixed-yield model. It means that if $rank(A_1)=m$ and $m\leq n$, we can find a set of $P=[P_s]$ to satisfy $\sum_{s:(u,s)\in UO}P_s\cdot VS_s=P'\cdot VU_u=P'\cdot (VC_1+VC_2+...+VC_m)$ where $VU_u=\sum_{s:(u,s)\in UO}VS_s$, so that Equation\ref{Obj1} can be transferred into the following form:
\begin{equation}
\begin{split}
    Obj_1&=\sum_{s:(u,s)\in UO}(P_s\cdot VS_s +\sum_{q\in Q} P_{s,q}\cdot QS_{s,q})+P^{'}_{u}VU^2_{u}\ \ \ \ \forall u\in CDU \\
    &=(P'\cdot VU_u+VU_u^2)+\sum_{s:(u,s)\in UO}\sum_{q\in Q} P_{s,q}\cdot QS_{s,q}\ \ \ \ \forall u\in CDU
\end{split}
\end{equation}

From this equation, we can observe that when an appropriate $P_s$ is selected, the objective function consists of two main components: one is the total feed to the CDU, and the other is related to the quality of the fractions. By choosing suitable $P'$ and $P_u$, we can ensure that the feed rate meets the desired value. Additionally, since $QS$is solely dependent on the crude oil formulation, we can use the $P_{s,q}$ term to impose a penalty or reward on the fraction quality, thus adjusting the feed composition. This effectively explains the intrinsic meaning and rationale behind the objective function we propose.

If $rank(A_1)<m$, then we need to find whether $rank\begin{pmatrix}
A_1 \\
A_2 
\end{pmatrix}=m$ or $rank\begin{pmatrix}
A_1 \\
A_2 \\
A_3
\end{pmatrix}=m$, so that we modify the objective function into 
\begin{equation}
\begin{split}
    Obj_i=\sum_{s:(u,s)\in UO}(&P_s\cdot VS_s +\sum_{q\in SQV} P_{s,q}\cdot QS_{s,q}\cdot VS_{s}+\sum_{q\in Q} P_{s,q}\cdot QS_{s,q})\\
    &+P^{'}_{u}VU^2_{u}\ \ \ \ \ \ \ \ \ \ \ \ \ \ \ \ \ \ \ \ \  \ \ \ \ \ \ \ \ \ \ \ 
 \ \ \ \ \ \ \forall u\in CDU 
\end{split}
\end{equation}
or 
\begin{equation}
\begin{split}
    Obj_i=\sum_{s:(u,s)\in UO}(&P_s\cdot VS_s +\sum_{q\in SQV} P_{s,q}\cdot QS_{s,q}\cdot VS_{s}+\sum_{q\in Q} P_{s,q}\cdot QS_{s,q}\\
    &+\sum_{q\in SQW}P_{s,q}\cdot QS_{s,'sg'}\cdot VS_s)+P^{'}_{u}VU^2_{u}\ \ \ 
 \ \ \ \ \ \ \forall u\in CDU 
\end{split}
\end{equation}

It is worth noting that due to the inherent characteristics of the CDU model, including constraints and domain limitations, it is not necessary to strictly adhere to the aforementioned form of the objective function in practical applications. Through extensive experimentation, we have found that the following form of the objective function is sufficient to meet the vast majority of the requirements.

\begin{equation}
    Obj_{i,t}\ = \sum_{s:(u,s)\in UO}(P_{s,t}VS_{s,u,t}+\sum_{q\in Q}P_{s,q,t}\cdot QS_{s,q,t})\ \ \ \ \ \ \forall t\in T,u\in CDU
\end{equation}
or
\begin{equation}
    Obj_{i,t}\ = \sum_{s:(u,s)\in UO}(P_{s,t}VS_{s,u,t}+\sum_{q\in Q}P_{s,q,t}\cdot QS_{s,q,t})+P_{u,t}VU_{u,t}^2\ \ \ \ \ \ \forall t\in T,u\in CDU
\end{equation}

And the proof for storage tanks, since a storage tank only needs to store one type of intermediate product in our model, the objective function can be decomposed as follows:
\begin{equation}
    Obj_2 = \sum_{s:(u,s)\in IO} P'_{s,t}\cdot VI_{s,i,t}
\end{equation}
\begin{equation}
    P'_{s,t}=P_{s,t}-\alpha I_{i,t}
\end{equation}
Combining the above two equations together,we can get that:
\begin{equation}
    Obj_2 = \sum_{s:(u,s)\in IO}(P_{s,t}-\alpha I_{i,t})VI_{s,i,t}
\end{equation}
\begin{equation}
    I_{i,t}+\sum_{s':(i,s')\in IO}VI_{s,i,t} = I_{i,t-1}+\sum_{s:(i,s)\in II}VI_{s,i,t}
\end{equation}
Since we solve the refinery planning problem in a sequential manner, we first define the objective function of the CDU and obtain the values of its decision variables through mathematical programming, thereby determining the feed flow and quality of the storage tanks. Therefore, when solving the storage tanks sub model, the total quantity and quality of the existing contents in the storage tanks on the right-hand side of the equation $Eq(A.)$ are determined and denoted as $I^{total}_{i,t}$.
\begin{equation}
    I_{i,t}+VI_{s,i,t}=I_{i,t}^{total}
\end{equation}
\begin{equation}
    Obj_2 = \sum_{s:(u,s)\in IO}(P_{s,t}-\alpha I_{i,t})(I_i^{total}-I_{i,t})
\end{equation}
which become a quadratic function with respect to the inventory level. Therefore, the agent provides different $P_{s,t}$ values based on varying external conditions to regulate the inventory level.

It is worth noting that, similar to CDU, not every intermediate product's inventory needs to be defined. Particularly due to the presence of nonlinear recipes, if the inventory of each product requires adjustment by the agent, infeasible solutions are likely to arise. The suggestion provided in this study is to only specify the inventory of certain key intermediate products, such as LN, KERO and DS, in order to ensure feasible solutions.

The remaining part of the refinery including secondary unit and blenders, can also be decomposed or integrated into one. As the external market give a strict quality requirement for the product, it is hard to find an alternative simple model is challenging. Here, we take the profit function itself as the objective function of the sub model and use objective consistency to ensure that the proposed method's potential is no lower than that of mathematical programming methods. It is evident that the optimal solution $\omega^*$ must inevitably be a feasible solution within the feasible domain. And we have previously proven that the sub model of CDU and storage tanks can achieve any desired value within the feasible domain, as long as the parameters are properly set. When the CDU and storage tanks get theirs optimal solution $\omega_1^*$ and $\omega_2^*$, the input flow to sub model 3 are consistent with that of the optimal solution of mathematical programming, so that $\omega_3^*$ must be a feasible and optimal solution for the sub model 3 as we haven't modify its constraints. 

Finally, the solutions of sub models got by sequence $\omega_1^*$,$\omega_2^*$ and $\omega_3^*$ come into being the optimal planning strategy for refinery.

\end{document}